\documentclass[twocolumn,amssymb,footinbib, nobibnotes, aps, reprint]{revtex4-1}

\providecommand{\floor}[1]{\left \lfloor #1 \right \rfloor}

\usepackage{todonotes}
\usepackage{bm}
\usepackage{graphicx}
\usepackage{amsthm,amssymb}
\usepackage{amsmath}
\usepackage{datetime}
\usepackage{braket} 
\usepackage{lipsum} 

\usepackage{hyperref}
\hypersetup{colorlinks,
	urlcolor=[rgb]{0,0.6,0.6},
	linkcolor=[rgb]{0,0.0,0.8}}

\usepackage{cleveref} 

\graphicspath{{figures/}}

\begin{document}

\title{Phase space analysis of quantum transport in electronic nanodevices}

\author{George Datseris}
\email{george.datseris@ds.mpg.de}
\affiliation{Max Planck Institute for Dynamics and Self-Organization}
\affiliation{Institute for the Dynamics of Complex Systems, Georg-August-Universit\"at G\"ottingen}
\author{Ragnar Fleischmann}
\affiliation{Max Planck Institute for Dynamics and Self-Organization}
\affiliation{Institute for the Dynamics of Complex Systems, Georg-August-Universit\"at G\"ottingen}
\date{\today}

\begin{abstract}
Electronic transport in nanodevices is commonly studied theoretically and numerically within the Landauer-B\"uttiker formalism: a device is characterized by its scattering properties to and from reservoirs connected by perfect semi-infinite leads, and transport quantities are derived from the scattering matrix. In some respects, however, the device becomes a ``black box'' as one only analyses what goes in and out. Here we use the Husimi function as a complementary tool for quantitatively understanding transport in graphene nanodevices. It is a phase space representation of the scattering wavefunctions that allows to link the scattering matrix to a more semiclassical and intuitive description and gain additional insight in to the transport process. 
In this article we use the Husimi function to analyze some of the fascinating electronic transport properties of graphene, \emph{Klein tunneling} and \emph{intervalley scattering}, in two exemplary graphene nanodevices.
By this we demonstrate the usefulness of the Husimi function in electronic nanodevices and present novel results e.g. on Klein tunneling outside the Dirac regime and intervalley scattering at a pn-junction and a tilted graphene edge.
\end{abstract}

\maketitle

\section*{Introduction}
\begin{figure*}[t!]
	\includegraphics[width=\textwidth]{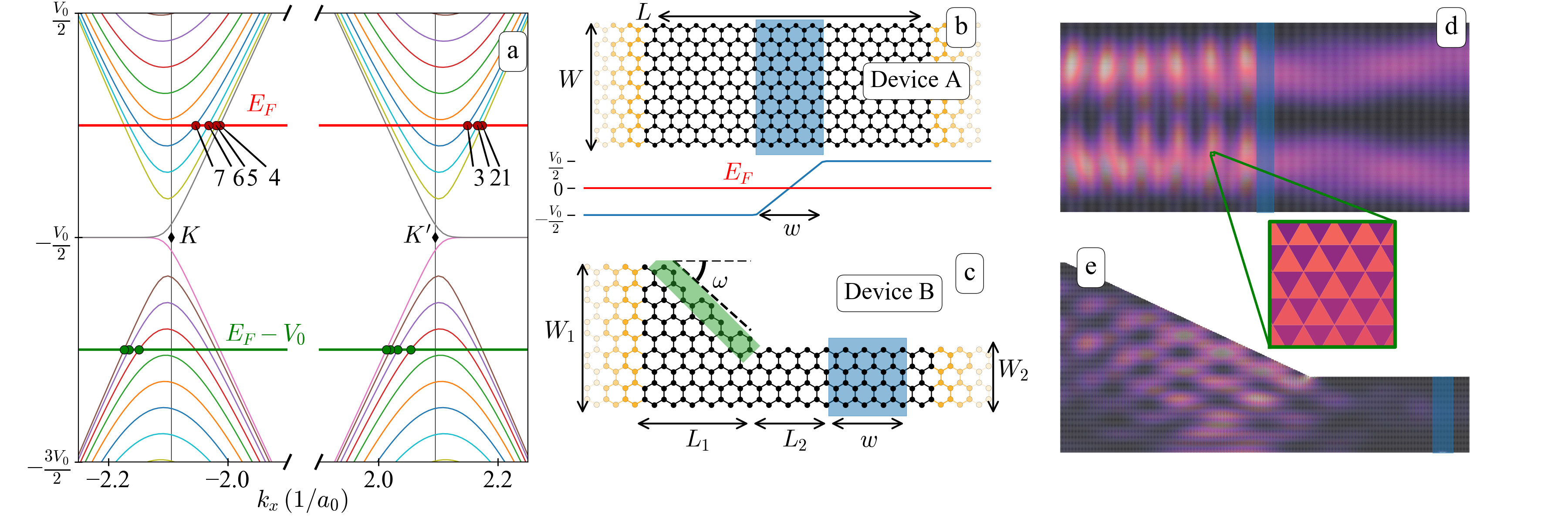}
	\caption{Graphene nanodevices. (a) Dispersion relation of a zigzag graphene nanoribbon, which is separated into two inequivalent Dirac valleys for small energies. 
		Sketched are the levels of the incoming (red) and outgoing (green) energy as well as the incoming and outgoing modes (intersections with the bands).
		As our transport setup is from the left to the right lead, only modes with positive group velocity (slope at the intersection) are valid.
		Incoming mode numbers are also shown.
		(b) Sketch of device A, a simple graphene nanoribbon with a pn-junction (blue). Below the device we sketch the potential profile of the pn-junction. 
		(c) Sketch of device B. With green we highlight the scattering edge. Leads are colored orange in both sketches. 
		(d, e) A scattering wavefunction amplitude inside simulated devices A and B respectively. The inset is showing how we plot the wavefunctions: for each sublattice we use a different marker (up or down triangles).}
	\label{fig:devices}
\end{figure*}

Typically quantum transport simulations of electronic nanodevices are based on the Landauer-B\"uttiker formalism. 
There, a nanodevice is regarded as a \textit{scattering region} that is connected to electron reservoirs by semi-infinite leads (see Fig.~\ref{fig:devices}(b, c) for examples).
The central quantity of this formalism is the scattering matrix $S$, from which one can obtain transmission probabilities, electric and thermal conductivities, as well as other useful quantities~\cite{Datta1995}. Even though this approach gives a wealth of information on transport through the device, in some respects the device appears to be a ``black box''.

This becomes an apparent weakness when one wants to connect the quantitative results the formalism produces to the physical intuition obtained by the semiclassical picture, or when trying to understand the role played by the different components of a complex (not easy to compartmentalize) device.
In that case one wants to analyse the scattering wavefunctions inside the device, and how they populate position and momentum space.
For example, if one wants to study and understand Klein tunneling (briefly reviewed in sec.~\ref{sec:kleintheory}) in a graphene nanodevice, then information about the momentum orientation (``angle of incidence'') of the wavefunction inside the device (and specifically before a pn-junction) is important.
In order to complement the scattering matrix information, and to get an intuitive connection with the semiclassical picture, here we will use the \emph{Husimi function} $Q$ which transforms a wavefunction into a phase space (quasi-)distribution.

Husimi functions have been introduced to quantum mechanics a long time ago~\cite{Husimi1940} and since then have been used in various areas of physics, like quantum optics~\cite{Schleich2001,Moya-Cessa2008} and ocean acoustics~\cite{Virovlyansky2012}. Husimi functions also play a prominent role in the field of quantum chaos, which tries to unravel the properties of complex quantum systems. 
For example, they have been used to understand the structure of the eigenfunctions in paradigmatic chaotic systems like quantum maps and billiards~\cite{Nonnenmacher1998,Baecker2003,Baecker2004,Baecker2005,Toscano2008}, transport in quantum ratchets~\cite{Schanz2005}, the properties of optical microdisc lasers~\cite{Hentschel2003,Wiersig2008,Baecker2009} and even electronic transport in disordered systems~\cite{Feist2006}. Recently Mason et al.~\cite{Mason2013a, Mason2013, Mason2015} have introduced a \emph{processed Husimi map} in tight-binding models of nanodevices allowing to recover and visualize classical paths in coordinate space. But in general solid state physics does not yet take much advantage of this useful tool.

In this article we will use the Husimi function to analyse graphene nanodevices.
Graphene is a fascinating material for studying quantum transport, due to the abundance of new physics it brought into light quickly after its discovery~\cite{DasSarma2011} e.g. weak (anti-)localization effects connected with the existence of two inequivalent valleys~\cite{McCann2006} or the Klein tunnel effect and its potential impact on technological applications~\cite{Katsnelson2006}.
Motivated by the intention to understand the impact of \emph{Klein tunneling} (sec.~\ref{sec:kleintheory}) and \emph{intervalley scattering} on transport in arbitrary graphene nanodevices, in this article we will use the Husimi function as a distribution in position and momentum space to analyse two exemplary devices. 
We find that the the combination of Husimi functions and semiclassical considerations is a powerful tool to understand transport phenomena in graphene nanodevices.
The observations we report also include the mode dependence of intervalley scattering at pn-junctions and the behaviour of Klein tunnelling at trigonal warping (sec.~\ref{sec:deviceA}) as well as the quantification of intervalley scattering at a tilted graphene edge (sec.~\ref{sec:deviceB}).

\newpage

\section{Theory}

\subsection{Klein tunneling in graphene}
\label{sec:kleintheory}

Klein tunneling is a paradox of relativistic quantum mechanics first discovered by Klein in 1929~\cite{Klein1929,Calogeracos1999}, but it has become a reality in graphene~\cite{Katsnelson2006}. This is due to the fact that low-energy excitations in graphene are well described by the Dirac equation for massless relativistic particles with linear energy dispersion, $\epsilon \sim |\mathbf{k}|$, and the wave function on the honey comb lattice of graphene, separated in two sublattices, is represented by a spinor, corresponding to the Dirac spinor.  In this formalism \textit{chirality-conservation} leads to a fascinating transport anomaly (for more details and a pedagogical review see Ref.~\cite{Allain2011}): conduction electrons are able to penetrate (almost) arbitrarily high potential barriers! If a conduction electron impinges normally on a potential step higher than the Fermi energy it will not get reflected but fully transmitted into a symmetric state in the valence band inside the high potential region instead.

More quantitatively, consider a plane wave (an electronic eigenstate of the Dirac equation~\cite{Allain2011}) with wavevector $\mathbf{k}$ incident on a pn-junction (i.e.\ a potential step), like the one of Fig.~\ref{fig:devices}b. 
The transmission probability depends on the angle of the wavevector (the \textit{angle of incidence}) $\phi_\text{in} = \arctan(k_y/k_x)$ and on the width of the junction $w$~\cite{Katsnelson2006, Cheianov2006}, where for a very steep junction ($w\rightarrow0$) one expects 
\begin{equation}
T_{\text{Step}}(\phi_\text{in})  = - \frac{\cos(\phi_\text{in})\cos(\phi_\text{out})}{\sin^2\left(\tfrac{1}{2}(\phi_\text{in} + \phi_\text{out} )\right)} \label{eq:Tstep} 
\end{equation}
and for wide (smooth) junctions ($w \gg \lambda_F$)
\begin{equation}
T_\text{WKB}(\phi_\text{in}) = \exp\left(-\pi\frac{2k_1^2}{k_1 + k_2}\frac{w}{2}\sin^2\phi_\text{in}\right) \label{eq:TWKB}.
\end{equation}
Here $\lambda_F$ is the Fermi wavelength of the electrons, $\phi_\text{out}$  the angle of the wavevector of the transmitted wave (the \textit{angle of refraction}) and $k_{1,2}$ are the wavenumbers of the incident and the transmitted wave, respectively. In both expressions, at normal incidence, i.e.\ for $\phi_\text{in} = 0$, the transmission is perfect, $T=1$, independent of the height and shape of the potential step! 

The transmission probability in the Klein tunneling process crucially depends on the angle of incidence of the electron on the junction. This angle is clearly defined for a plane wave: the wavevector angle. But what is the ``angle of incidence'' of a more complex electron wavefunction in a realistic scenario, like the scattering wavefunction in a nanodevice of arbitrary spatial complexity? Basic examples of such wavefunctions are shown in Fig.~\ref{fig:devices}(d, e). In general a plane wave will not always be a good representation of these wavefunctions. Yet, if one wants to understand the impact of Klein tunneling on transport in such nanodevices, one  needs information about an angle of incidence. Because such information is inherently present in the (classical) phase space, below we will use the Husimi function to obtain this information. But first let us introduce the nanodevices we will be using as examples. 

\subsection{Graphene devices}
\label{sec:devices}
\label{sec:gnr_theory}
In our work we will study transport in tight-binding models of the two graphene-based devices shown in Fig.~\ref{fig:devices}.
Notice however that our methodology can be applied to tight-binding based quantum transport in any device (and in 3 dimensions just a well), as it will become clear below.
For the graphene devices, Device A is the conceptually simplest device in which one can study Klein tunneling in a realistic scenario (i.e. a finite nanodevice): a graphene nanoribbon (GNR) of constant width with a pn-junction in its middle.
In device A the boundary conditions are chosen such that it forms a ``zigzag'' nanoribbon. 
These have been studied by Bray and Fertig in detail within the Dirac approximation~\cite{Brey2006a} and many of their properties are known analytically (for small Fermi energies). 
Analytical descriptions in this case are possible because of the many symmetries that are present. Device B however breaks the conservation of $k_y$ as well as the reflection symmetry along the $x$ axis.
Note also that for $\omega = \pi/6$ the ``scattering edge'' in device B (highlighted in green in Fig.~\ref{fig:devices}c) exactly is an armchair boundary. 
In both devices we create pn-junctions via a linear increment of the potential energy from the n region with bias voltage $-V_0/2$ to the p region with bias $+V_0/2$ over a range $w$ (see Fig.~\ref{fig:devices}). The \emph{kinetic energy} $E$ of the incoming electrons is connected to the Fermi energy $E_F$ by $E = E_F + V_0/2$. 

Zigzag GNRs have a dispersion relation shown in Fig.~\ref{fig:devices}a and discussed in detail in Refs.~\cite{Brey2006a, Akhmerov2008}. For a given Fermi energy $M$ bands of the dispersion intersect the energy level at positive slope, thus having positive group velocity. This results in $M$ \emph{incoming} modes ($M$ is always odd and scales linearly with the width of the GNR). We order the modes by decreasing $k_x$, as shown in Fig.~\ref{fig:devices}a. Importantly, for small energies the two (inequivalent) Dirac valleys $K, K'$ are well separated in momentum space, which leads to the incoming modes being \emph{valley-polarized}. This means that modes 1 to $\floor{M/2}$ (where $\floor{\cdot}$  denotes the integer part) come from valley $K'$, while modes $\floor{M/2}+1$ to $M$ come from valley $K$. We also stress that $K$ has one \emph{additional} incoming mode, see Fig.~\ref{fig:devices}a.

All of our quantum transport simulations are tight binding calculations performed with the software Kwant~\cite{Groth2014}. The devices are finite \emph{scattering regions} that are coupled to semi-infinite \emph{leads} (which are also GNR). 
The \emph{modes} (eigenfunctions) of the leads enter the device and are subsequently scattered, defining the scattering wavefunctions $\psi_m$ for each mode. 
As we will consider transport always from the left to the right lead, we will only use a part of the scattering matrix which we define as  the $N\times M$ transmission matrix $\mathcal{T}$, where  $M$ and $N$ are the total number of modes in the left and right lead, respectively. 
The element $\mathcal{T}_{nm}$ is the transmission amplitude from the $m$-th (incoming) mode of the left  lead to the $n$-th (outgoing) mode of the right lead.
The total transmission probability $T_m$ of each individual incoming mode is given by
\begin{equation}
T_m = \sum_{i=1}^{N} \left| \mathcal{T}_{im}\right|^2\,.
\label{eq:Tm}
\end{equation}

\subsection{The Husimi function}
\label{sec:husimi_def}

In this section we define the Husimi function, which transforms a wavefunction into a phase space quasi-probability distribution. Let $\Ket{\mathcal{W}(\mathbf{r}_0,\mathbf{k}_0, \sigma)}$ denote a Gaussian wavepacket. In position representation and in the absence of magnetic fields this is simply~\cite{Takahashi1986}
\begin{equation}
\mathcal{W}(\mathbf{r}; \mathbf{r}_0,\mathbf{k}_0, \sigma) =	N_\sigma^{D/2} \exp\left( -\frac{\delta\mathbf{r}^2}{4\sigma^2} + i\,\mathbf{k}_0 \cdot \mathbf{r}\right)
\label{eq:wavepacket}
\end{equation}
(with $\delta\mathbf{r} = \mathbf{r} - \mathbf{r}_0$ and $D$ spatial dimensions)
which is a Gaussian envelope in space with origin $\mathbf{r}_0$ multiplying a plane wave with wavevector $\mathbf{k}_0$. $N_\sigma = \left(\sigma\sqrt{2\pi}\right)^{-1}$ is the normalization factor in the case of continuous space, so that $\Braket{W|W}=1$ and that $\Delta x = \Delta y = \sigma$. The key property of these wavepackets is that they minimize the uncertainty relation between position and momentum. Here $\sigma$ is the spatial uncertainty and thus is a parameter that controls the trade-off between the uncertainty in position ($\sigma$)  or momentum space ($1/(2\sigma)$). 

The Husimi function $Q$ is defined as the magnitude of a \emph{projection} of a wavefunction onto $\Ket{\mathcal{W}}$~\cite{Husimi1940, Harriman1988, Takahashi1986, Heller2018}
\begin{equation}
Q[\psi](\mathbf{r}_0,\mathbf{k}_0;\sigma)  = \frac{1}{\pi}\left| \Braket{\psi | \mathcal{W}(\mathbf{r}_0,\mathbf{k}_0;\sigma)} \right|^2
\label{eq:husimi_def}
\end{equation}
where for continuous space systems we have
\begin{equation}
\Braket{\psi | \mathcal{W}(\mathbf{r}_0,\mathbf{k}_0;\sigma)} =
\int \psi^*(\mathbf{r})  \times \mathcal{W}(\mathbf{r}, \mathbf{r}_0,\mathbf{k}_0; \sigma) \, d\mathbf{r}
\label{eq:husimi_def_continuous} 
\end{equation}
where the integration extends over the full spatial domain of the device (in our case in two dimensions). 
For a tight-binding system the projection is turned into a sum due to the discrete nature of the lattice
\begin{equation}
\Braket{\psi | \mathcal{W}(\mathbf{r}_0,\mathbf{k}_0,\sigma)} = \sum_j \psi^*(\mathbf{r}_j)\times e^{-\frac{\delta \mathbf{r}_j ^2}{4\sigma^2}}e^{i\mathbf{k}_0 \cdot \mathbf{r}_j}
\label{eq:husimi_def_discrete}
\end{equation} 
with $\delta \mathbf{r}_j = \mathbf{r}_j- \mathbf{r}_0$, $\psi(\mathbf{r}_j) \equiv \psi_j$ being the wavefunction at lattice site $j$ with position $\mathbf{r}_j$~\footnote{For each $\mathbf{r}_0$ we use only lattice sites that are within $|\mathbf{r}_j - \mathbf{r}_0|\leq 3\sigma$, to reduce computation time. Notice the complex conjugation $\psi^*$ in eqs.~\eqref{eq:husimi_def_continuous} and \eqref{eq:husimi_def_discrete}, which sometimes is omitted in the literature.  While for closed systems with time-reversal symmetry it can be omitted, it is crucial for open systems, which we explore here, and for systems with broken time reversal symmetry like those in magnetic fields.}. The normalization factors here depend on the lattice and thus in the following we will omit normalizations. $Q$ is a rigorous method for transforming a wavefunction into a phase space distribution (Weierstrass transform of the Wigner function~\cite{Takahashi1986}) and is used as a versatile tool for understanding complex quantum and other wave dynamics~\cite{Virovlyansky2012,Nonnenmacher1998,Baecker2003,Baecker2004,Baecker2005,Toscano2008,Schanz2005,Hentschel2003,Wiersig2008,Baecker2009,Feist2006,Mahmud2005,Mason2013,Mason2013a,Mason2015,Takahashi1986, Harriman1988, Heller2018}.

\subsection{Husimi function in a lead} 
\label{sec:lead_Q}
Before moving on to the numeric applications of this paper, it is helpful to obtain some intuition about $Q$ in analytically treatable examples. Let us first simply consider a plane wave $P(\mathbf{r}, \mathbf{k}) = \exp(i\mathbf{k}\cdot\mathbf{r})$. It is straightforward to calculate its Husimi function with
\begin{equation*}
\Braket{P | \mathcal{W}(\mathbf{r}_0,\mathbf{k}_0,\sigma)} =  
4\pi\sigma^2 e^{-k_x'^2\sigma^2}e^{-k_y'^2\sigma^2}e^{ik_x'x_0}e^{ik_y' y_0}, \end{equation*}
where we used
\begin{equation*}
\int_{-\infty}^{+\infty} e^{-\frac{(x-x_0)^2}{4\sigma^2}}e^{ikx}\, dx = 2\sigma \sqrt{\pi}e^{-k^2\sigma^2}e^{ikx_0}.
\end{equation*}
We thus find
\begin{equation}
Q[P](\mathbf{r}_0,\mathbf{k}_0;\sigma) \propto e^{-2k_x'^2\sigma^2}e^{-2k_y'^2\sigma^2}
\end{equation}
with $k' = k - k_0$. As one would expect the Husimi function $Q$ does not depend on $\mathbf{r}_0$, since a plane wave is spatially homogenous, and $Q$ is a Gaussian in momentum space since the integral above is the Fourier transform of a Gaussian. In addition, $Q$ only depends on the difference between the wavevectors of the Gaussian wave packet and the plane wave.

Let us now examine the case of a lead (or waveguide). Here the wavefunction $X$ is a plane wave in the longitudinal direction, while the transverse part is the quantum well wavefunction, i.e.
\begin{equation}
X = e^{ik_x x}\sin(k_m y).
\end{equation}
Notice that the following analytic result applies to both square lattice leads (whose low-energy excitations follow a dispersion $\epsilon\sim k^2$ and we know explicitly $k_m = m\pi/W$ for a lead of width $W$), but also zigzag graphene nanoribbons (GNRs) because the transverse wavefunctions there are as well sine modes (see end of sec.~\ref{sec:gnr_theory}). The only difference is that for GNRs the expression for $k_m$ is given by eq.~\eqref{eq:km_for_gnr}.

As we know that in the Husimi function the longitudinal component will result to $Q_x = e^{-2k_x'^2\sigma^2}$, we now tend to the integral of the sine function with the Gaussian function in finite limits. Although analytically solvable, its expression is not so trivial

\begin{figure}[t!]
	\includegraphics[width=\columnwidth]{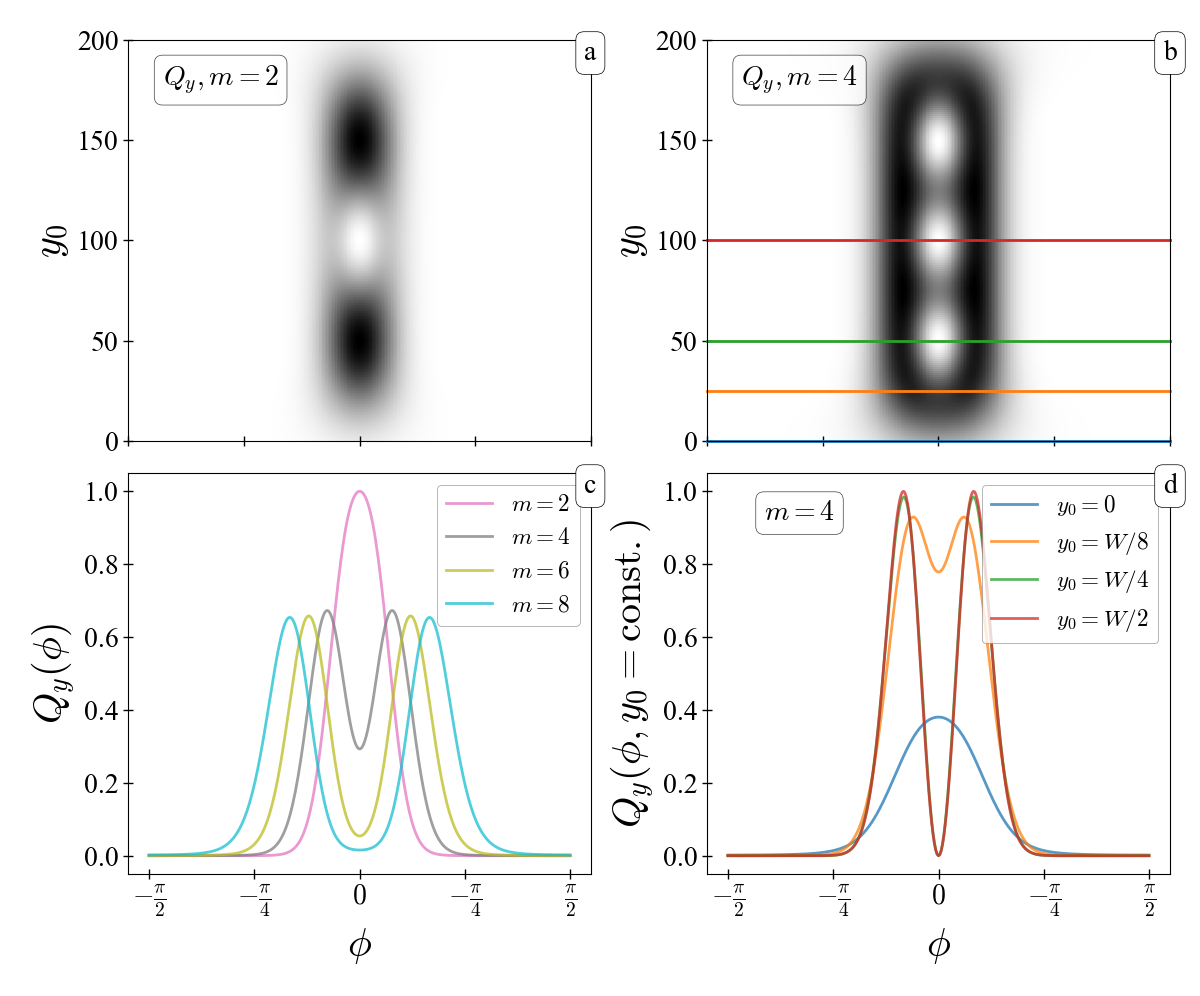}
	\caption{
		Equation~\eqref{eq:lead_Q} for the case of $k_m=m\pi/W$, with $W=200,\sigma=16,k_F=0.25$ (chosen semi-arbitrarily).
		(a, b) $Q_y$ for different $m$. (c) Marginal distribution $Q_y(\phi)$ (eq.~\eqref{eq:marginal_analytic}) for different $m$.
		(d) Cuts through $Q_y$ for different $y_0$ (also shown in panel (b)).
	}
	\label{fig:lead_Q}
\end{figure}

\begin{widetext}
	\begin{align}
	Q_y = \left| \int_{0}^{W} \sin\left(\frac{m\pi}{W}y\right) \exp\left(\frac{(y - y_0)^2}{4\sigma ^2} + i k_{y,0}y \right) d\!y \right|^2 = 
	\frac{1}{4} \pi  \sigma ^2 \exp\left(-2(k_m + k_{y,0})^2\sigma^2\right) \times 
	\nonumber \\ 
	\left|
	e^{4 k_m k_{y,0} \sigma ^2} \left(\text{erf}\left(\frac{W-y_0}{2 \sigma }+i \sigma  (k_m-k_{y,0})\right)+\text{erf}\left(\frac{y_0}{2 \sigma }-i \sigma  (k_m-k_{y,0})\right)\right)
	+ \right. 
	\nonumber \\
	\left.
	e^{2 i k_m y_0} \left(\text{erf}\left(\frac{y_0-W}{2 \sigma }+i \sigma  (k_m+k_{y,0})\right)-\text{erf}\left(\frac{y_0}{2 \sigma }+i \sigma  (k_m+k_{y,0})\right)\right)
	\right|^2
	\label{eq:lead_Q}
	\end{align}
\end{widetext}
with $\text{erf}$ the error function, and keeping in mind that $Q[X] = Q_x Q_y$. 
In our later analysis we will use $k_{y,0} = k_F\sin(\phi)$, with an appropriate $k_F$, as discussed in further detail in sec.~\ref{sec:calc_Q}. To obtain an intuition around Eq.~\eqref{eq:lead_Q} we are plotting it in Fig.~\ref{fig:lead_Q}. Notice that $Q_y = Q_y(\phi, y_0)$, i.e. it is a function of two quantities ($\phi$ because $k_{y,0}=k_F \sin(\phi)$). Based on this $Q_y$ we can define a marginal distribution over $\phi$ as
\begin{align}
Q_y(\phi) = \int_{0}^{W}Q_y(\phi, y_0) d\!y_0.
\label{eq:marginal_analytic}
\end{align}

Here there are two properties to point out. $Q$ is symmetric over $k_m = 0$ (and equivalently $\phi = 0$). This is already true from eq.~\eqref{eq:lead_Q}, but it can also be understood simply from the fact that $\sin(k_m y)$ decomposes to $(e^{ik_m y}-e^{-ik_m y})/2i$, which is a superposition of two plane waves with opposite directions. Furthermore, as $y_0$ moves further away from the center of the lead $W/2$, the accuracy of the Husimi function drops significantly (Fig.~\ref{fig:lead_Q}(d)). This has implications for calculating Husimi functions exactly at the edges of a tight-binding device, which we will discuss in sec.~\ref{sec:intervalley}.

\subsection{Calculating the Husimi function}
\label{sec:calc_Q}

For each scattering wavefunction $\psi_m$ we compute the Husimi function $Q$.
To reduce the dimensionality of (the parameters of) $Q$ we only evaluate it at well chosen transverse cuts ($x_0=const.$), e.g.\ just in front or behind the pn-junction. We will thus obtain a distribution of incoming and outgoing wavevectors that ``pass through'' these cuts as a function of the transverse coordinate $y_0$. 

\begin{figure*}[t!]
	\includegraphics[width=\textwidth]{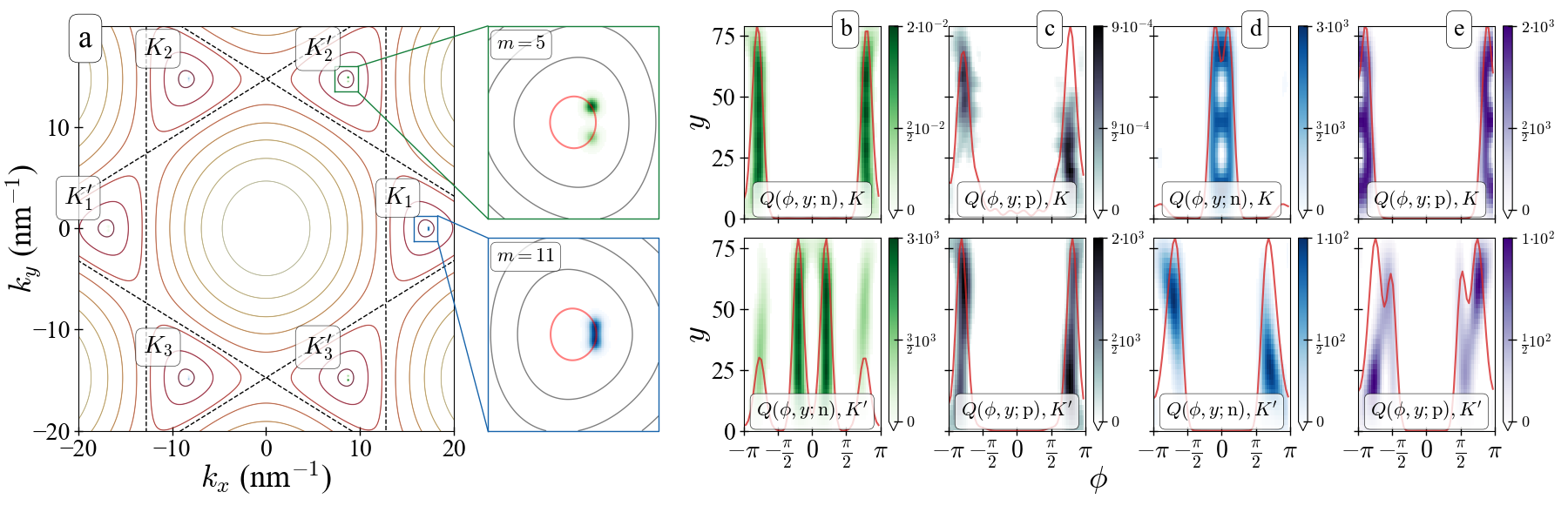}
	\caption{Husimi functions in device A. (a) Dispersion relation of graphene (red-yellow color, dashed black line for $E = t = 2.8$ eV~\cite{CastroNeto2009}, see~\cite{Goerbig2011a} for a derivation) and Husimi functions for 2 incoming modes (blue and green respectively) over the entire Brillouin zone, for $\mathbf{r}_0 = (\frac{L}{2}, \frac{W}{2}), E_F = 0.2\, \text{eV}, V_0 = 0\, \text{eV}, W = 80$ nm leading to $M = 17$ incoming modes. With $K_{i}^{(')}$ we label the six valleys. In the insets, the bright-red colored contour is noting the incoming energy of the simulation, here 0.2 eV. (b) Husimi distribution functions over wavevector angles $\phi$ and positions $y$ in device A, $3\sigma\approx 24$ nm before the pn-junction (i.e. incoming \& reflected) for valleys $K, K'$ (averaging over subscripts 1-3), see sec.~\ref{sec:calc_Q}. Here $V_0 = 0.4, E_F = 0.0$ eV (i.e. same incoming energy as panel (a)). (c) Same as (b) but $Q$ is measured $3\sigma\approx 24$ nm after the pn-junction (i.e. transmitted). For (b, c) the mode number is $m=5$. (d, e) Same as (b, c) but for mode number $m = 11$. Over all $Q(\phi, y)$ we plot the marginal distribution $Q(\phi)$ eq.~\eqref{eq:marginal} (averaged only over the three appropriate valleys instead of all six) with red color.}
	\label{fig:husimis}
	\label{fig:dispersion2D}
\end{figure*}

We further reduce the dimensionality of $Q$ by exploiting energy conservation. 
In Fig.~\ref{fig:husimis}a we show the two-dimensional dispersion of graphene
\begin{align}
\label{eq:dispersion}
\epsilon_\lambda(\mathbf{k}) &= \lambda t \sqrt{3 + f(\mathbf{k}} ) \\ 
f(\mathbf{k}) &= 2\cos\left(\sqrt{3}k_xa\right) + 4 \cos\left(\frac{\sqrt{3}}{2}k_xa\right)\cos\left(\frac{3}{2}k_ya\right) \nonumber 
\end{align}
with $\lambda =\pm1 $ the band index, $t=2.8$eV the hopping constant and $a\approx 0.142$nm the carbon-carbon distance. 
Measuring the Husimi function we see that $Q$ localizes strongly on the two-dimensional energy contour corresponding to the Fermi energy (even though in reality the dispersion relation of a finite GNR is in fact one-dimensional).
This allows us to reduce the dimensionality of $Q$ by using the 2D dispersion.
In the following we  measure $Q$ for wavevectors that populate the Fermi energy contour at equally spaced angles $\phi$. 

A difference with the honeycomb lattice versus the square lattice is that there are six valleys in the two-dimensional dispersion, as seen in Fig.~\ref{fig:dispersion2D}.
Thus we compute $Q$ in all six of them, and the angle is measured with respect to the respective Dirac point $K_\xi$ of each valley, i.e.\ $\phi=\arctan k'_y/k'_x$ with $\mathbf{k'}=\mathbf{k}-\mathbf{K_\xi}$. 
We denote this by $Q(\ldots;\xi)$, where $\xi$ counts the valleys ($\xi \in \{1,2 , 3\}$ means $K$, $\xi \in \{4, 5, 6\}$ means $K'$). 
This reduces $Q$ from depending on both $k_x, k_y$ to be only a function of $\phi$. 
The parameter $\sigma$ we will choose such that the wavevector uncertainty satisfies $\Delta k / k = \Delta \phi = 0.2$, where $k$ is the (average) magnitude of the wavevector with respect to Dirac point (see appendix~\ref{sec:angle_unc}).
This yields typical values of $\sigma \approx 8$ nm for small energies, while for higher energies $\sigma$ can be smaller than $4$nm. In the following and for device A the notation $x = \text{n}$ will denote a cut in the n region of the device, $3\sigma$ before the pn-junction, while $x = \text{p}$ will denote a cut $3\sigma$ after the junction. For device B the slice location is given explicitly (in the rest of the text we measure space in nm and energy in eV).

\begin{figure}[t!]
	\includegraphics[width=\columnwidth]{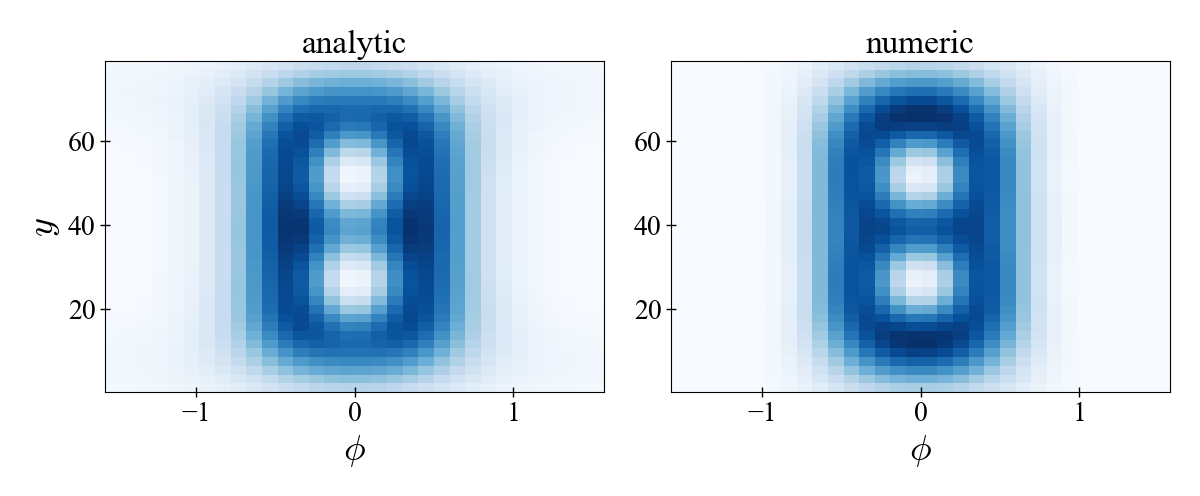}
	\caption{Eq.~\eqref{eq:lead_Q} using eq.~\eqref{eq:km_for_gnr} vs. numeric computation from $m=11$ (right panel is the same as panel (d) of Fig.~\ref{fig:husimis}).
	}
	\label{fig:analytic_vs_numeric}
\end{figure}

For the zigzag nanoribbon we can compare the numerically computed $Q$ with the analytical expression~\eqref{eq:lead_Q} (because we keep $x$ fixed, we in principle calculate $Q_y$ with our numerical scheme). 
The theory of Brey and Fertig shows that in the Dirac regime the transverse wavefunctions of a GNR are sine modes, $\sin(k_m y)$~\cite{Brey2006a} and thus their Husimi function is given by.~\eqref{eq:lead_Q}. To obtain the transverse wavevectors $k_m$ for zigzag GNRs one needs to solve~\cite{Brey2006a}
\begin{align}
\left(\frac{E}{\hbar v_F}\right)^2 - k_m^2 = \frac{k_m}{\tan(k_m W)}
\label{eq:km_for_gnr}
\end{align}
which cannot be solved analytically. We can thus find the $k_m$ either by solving Eq.~\ref{eq:km_for_gnr} numerically or by fitting sine functions to the transverse wavefunctions of the tight-binding simulations.
The analytical and numerical Husimi functions are shown in Fig.~\ref{fig:analytic_vs_numeric} and we find excellent agreement.

\section{Applications of the Husimi function: device A}
\label{sec:deviceA}
\subsection{Klein Tunneling}
\label{sec:klein}
We first want to test the usage of $Q$ in a well studied situation where much can be inferred analytically: Klein tunneling in device A at small energies~\cite{Brey2006a} (see \cite{Allain2011} for a review on Klein tunneling in graphene). Fig.~\ref{fig:husimis}(b-e) shows $Q(\phi, y)$ in device A for $W= 80,L=12\sigma\approx 96, E_F=0,V_0 = 0.4$. The top panels show $Q$ for valley $K_2$, the bottom for $K_2'$. We show $Q$ both before (incoming \& reflected) and after (outgoing) the pn-junction for two modes. What we have seen is that for device A before the junction, $Q$ in valley $K_2(')$ is the mirror reflection of $Q$ in valley $K_3(')$ while in valley $K_1(')$ we find an almost exact superposition of the $Q$s in $K_1(')$  and $K_2(')$.

Fig.~\ref{fig:dispersion2D} shows that for all modes the incoming $Q$ nicely localizes at a single angle. We also show in red the marginal distributions
\begin{align}
Q(\phi;\xi) &= \int_0^W Q(\phi, y;x,\xi)\,d\!y , \nonumber \\
 Q(\phi) & = \sum_{\xi} Q(\phi;x,\xi).
\label{eq:marginal} 
\end{align}
Because in this setup the incoming $Q$ is highly localized, we do not need the entire distribution and can simply choose the maximum location of $Q(\phi)$, $\Phi$, to represent the ``incoming angle'' for each mode
\begin{align}
\Phi = \text{argmax}\left[Q(\phi; x=\text{p})\right],\;\text{for}\;\phi\in \left[0, \frac{\pi}{2}\right) 
\label{eq:Phi}
\end{align}
(we use $Q$ of valley $K_2(')$ exclusively for this, and we also know which of the two valleys is the incoming one).

\begin{figure*}[t!]
	\includegraphics[width=\textwidth]{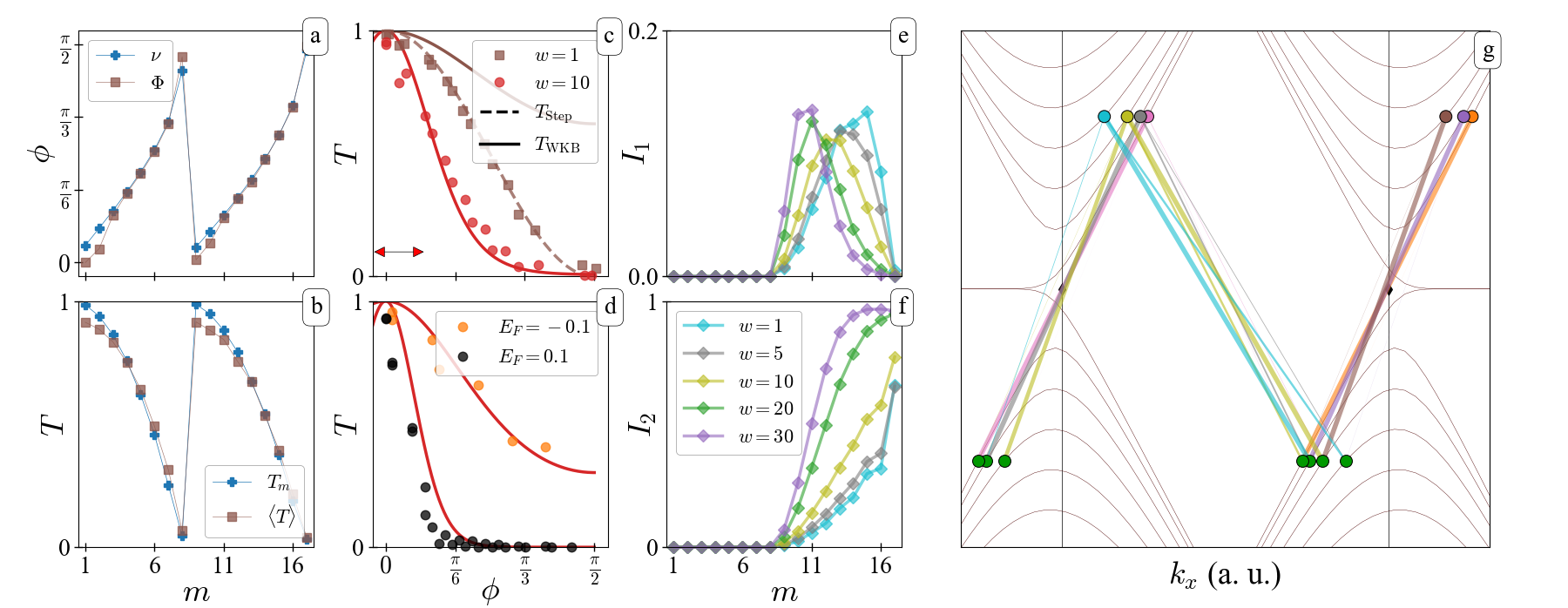}
	\caption{Klein tunneling and intervalley scattering in device A for $\sigma = 8, W = 80, L = 12\sigma, V_0 = 0.4, E_F = {0, \pm 0.1}$ and various $w$. (a) Angle of incidence $\nu$ deduced from the transverse wavefunctions, compared with the ones deduced from the Husimi function, $\Phi$. (b) Transmission probability obtained through the scattering matrix $T_m$ eq.~\eqref{eq:Tm} versus the integrated one obtained from $Q$, $\langle T \rangle$ using eq.~\eqref{eq:integrated}. (a, b) are plotted versus incoming mode $m$ and for $w=1$ nm. (c) Theoretical curves on Klein tunneling (lines, eqs.~\eqref{eq:TWKB}, \eqref{eq:Tstep}) and transmission probability $T_m$ versus $\Phi$ (scatter plots, for two different $w$ values). The red arrow notes the angle uncertainty $\Delta \phi$. (a, b, c) use $E_F = 0$ eV. (d) is the same as (c) but for $w=10$ and different Fermi energies instead.
	(e, f) Measures for intervalley scattering, eq.~\eqref{eq:intervalley} versus mode number. (g) Sketch (x-axis is not uniform) of where is each mode transmitted, based on the elements of the transmission matrix $\mathcal{T}$. The width of each line is proportionate to the transmission amplitude towards the outgoing channel that the line connects to (normalized to same maximum).}
	\label{fig:klein}
\end{figure*}

As discussed in the end of sec.~\ref{sec:calc_Q}, we can obtain numerically the wavevector of the transverse component of the wavefunction $k_m$, based on the theory of Bray and Fertig~\cite{Brey2006a} (notice that this is invalid outside the Dirac regime).
Kwant also provides $k_x$, the wavevector of the longitudinal component.
We then compare $\Phi$ with $\nu = \arctan(k_m / k_x)$ in Fig.~\ref{fig:klein}a. We see that only for the ``highest'' (meaning high energy band) modes of each cone $\Phi$ does not have a perfect agreement with $\nu$. We now want to use $\Phi$ to compare the results of the tight-binding calculations with theoretical result for the Klein tunneling at a pn-junction, eq.~\eqref{eq:TWKB} and~\eqref{eq:Tstep}. 
In Fig.~\ref{fig:klein}c we plot the theoretical curves and the values of $T_m$ versus $\Phi$ for each mode, for two different pn-junction widths $w$, and find very good agreement.
This does not only hold for the case of a symmetric pn-junction, i.e. $E_F=0$, but also for higher and lower Fermi energies, as shown in Fig.~\ref{fig:klein}d  for $w = 10$ nm (for other parameter values we also find excellent agreement).

It is clear that through $Q$ we can find the parameter $\phi_\text{in}$. Now we want to show that we can even obtain the transmission probabilities from the Husimi function using the theoretical transmission formulas. Using the marginal distribution of eq.~\eqref{eq:marginal} we can compute the transmission of a mode as the average 
\begin{align}
\langle T \rangle =\frac{ \int_{-\frac{\pi}{2}}^{\frac{\pi}{2}} T(\phi)Q(\phi; x=\text{p})\, d\phi}{\int_{-\frac{\pi}{2}}^{\frac{\pi}{2}} Q(\phi; x=\text{p})\, d\phi}
\label{eq:integrated}
\end{align}
where $T$ si either eq.~\eqref{eq:Tstep} or \eqref{eq:TWKB}. 
Notice that in principle eq.~\eqref{eq:integrated} could be resolved analytically, since we know the expressions for both $T$ as well as $Q$ (see eq.~\eqref{eq:lead_Q}) for the simple device A. Unfortunately, we were not able to indeed resolve the integral analytically, but numeric integration is always possible.

In Fig.~\ref{fig:klein}b we compare $\Braket{T}$ with $T_m$ and again we find a near perfect match (also for many more parameters than the ones shown). Equation~\eqref{eq:integrated} will also give a good estimate of the transmission value in cases where the distribution is not strongly localized at a single angle, allowing us to use the integrated transmission in more complicated cases like those in sec.~\ref{sec:deviceB}.

\subsection{Intervalley Scattering}
\label{sec:intervalley}
We now turn to study \emph{intervalley scattering}, which describes the scattering of a wavefunction from one valley to another (inequivalent) one, e.g. from $K$ to $K'$. We discussed in sec.~\ref{sec:gnr_theory} that for zigzag GNRs and low energies every incoming mode is valley-polarized~\cite{Brey2006a}.
Intervalley scattering has found considerable interest in the literature, and was first discussed in the context of weak localization~\cite{McCann2006, Morpurgo2006, Hilke2014, Yan2016}. Later work focused on valley filters and valley ``spintronics'', see~\cite{Rycerz2007, Gunlycke2011, Lee2017} and references therein. The discussions in the literature so far have been qualitative and mostly theoretical.

The Husimi function is an excellent tool to study intervalley scattering, because it directly provides information in momentum space at different positions in the device. In fact, Mason et al. have used a processed Husimi projection technique in Ref.~\cite{Mason2013} to study intervalley scattering in graphene billiards.  Here we will use a simpler approach directly using the Husimi function. As one can already see from Fig.~\ref{fig:husimis}b-e, the ``incoming $Q$'' (i.e. $Q(y,\phi)$ with $\phi \in \left[-\pi/2, \pi/2\right)$) has most weight in one valley (the ``incoming valley'') $V_i$, while the other (the ``complementary'') valley $V_c$ contains either just noise or only the reflected wave (compare the scales of the colorbars). In panels (c, e) it is evident there exist modes that undergo intervalley scattering, as for panel (e) the outgoing valley $K'$ has significantly more weight than what it had in the incoming case of panel (d).

We want to define two intuitive measures for intervalley scattering. We first define the following weights (the sums are over all equivalent valleys)
\begin{align}
\alpha & = \sum_{\xi  \in V_i} \int_{0}^{W}\!\! \int_{-\frac{\pi}{2}}^{\frac{\pi}{2}}Q(\phi, y;x=\text{n},\xi)\,d\phi\,dy \\
\beta & = \sum_{\xi \in V_i}\int_0^W\!\! \int_{-\pi}^{\pi}Q(\phi, y;x=\text{p},\xi)\,d\phi\,dy \\
\gamma & = \sum_{\xi \in V_c} \int_0^W \!\!\int_{-\pi}^{\pi}Q(\phi, y;x=\text{p},\xi)\,d\phi\,dy.
\end{align}
$\alpha$, is used for the normalization to the incoming mode. $\beta$ and $\gamma$ measure the weights of the transmitted wave that are localized in the same valley as the incoming mode and its complement, respectively. 
With these quantities we define
\begin{align}
I_1 = \frac{\gamma}{\alpha},\quad I_2 = \frac{\gamma}{\beta + \gamma}.
\label{eq:intervalley}
\end{align}
Here $I_1$ is the the fraction of the \textit{incoming wave} that is transmitted through the pn-junctions \textit{and} has undergone intervalley scattering. $I_2$ is the fraction of the \textit{transmitted wave} that has undergone intervalley scattering, i.e.~a transmitted wave with $I_2=0$ or $I_2=1$ is completely valley polarized. 
We show both measures of intervalley scattering in Fig.~\ref{fig:klein}e,f plotted versus the mode number for various junction widths. (Qualitatively the results remain unchanged when we use only $K_2(')$ instead of summing over equivalent valleys).

The most striking feature of Fig.~\ref{fig:klein}e,f is that intervalley scattering happens only for the second half of the modes. Recall that modes with $ 1 \le m \le \floor{M/2}$ come from $K'$ while the higher modes come from the $K$ valley which has an additional incoming band (see Fig.~\ref{fig:devices}a or Fig.~\ref{fig:klein}g). The perplexing result of Fig.~\ref{fig:klein}e can be qualitatively explained based on this extra mode and the unitarity of the scattering matrix $S$~\cite{Datta1995} (i.e. current conservation). To aid the following argument, in Fig.~\ref{fig:klein}g we show a sketch of where is each incoming mode transmitted. The lines connecting incoming and outgoing modes have widths directly proportional to the transmission amplitude $|\mathcal{T}_{im}|^2$.

After transmission, each mode ``tries'' to scatter into a the same valley at negative energy to conserve the \textit{valley pseudospin} (green dots in Fig.~\ref{fig:klein}g). Likewise should the reflected part scatter into modes in the same valley at the same energy level but with negative group velocity. Modes 1 to $\floor{M/2}$ have no problem achieving this, as within their valley the outgoing channels are more than the incoming ones and thus available channels always exist. This is not the case however for modes $\floor{M/2} + 1$ to $M$, since the number of outgoing channels \emph{within the same valley} is one less, both for transmission and reflection. As the mode number increases the outgoing channels are filled and the higher modes have to move some of their weight to other channels (as a specific outgoing channel cannot be filled with more than total transmission of 1, see Ref.~\cite{Datta1995}). The only remaining channels that can accommodate these modes exist in the $K'$ valley (right valley of Fig.~\ref{fig:devices}a) which leads to intervalley scattering.

\subsection{Trigonal Warping and Klein Tunneling}
\label{sec:higher_energy}

\begin{figure}[t!]
	\includegraphics[width=\columnwidth]{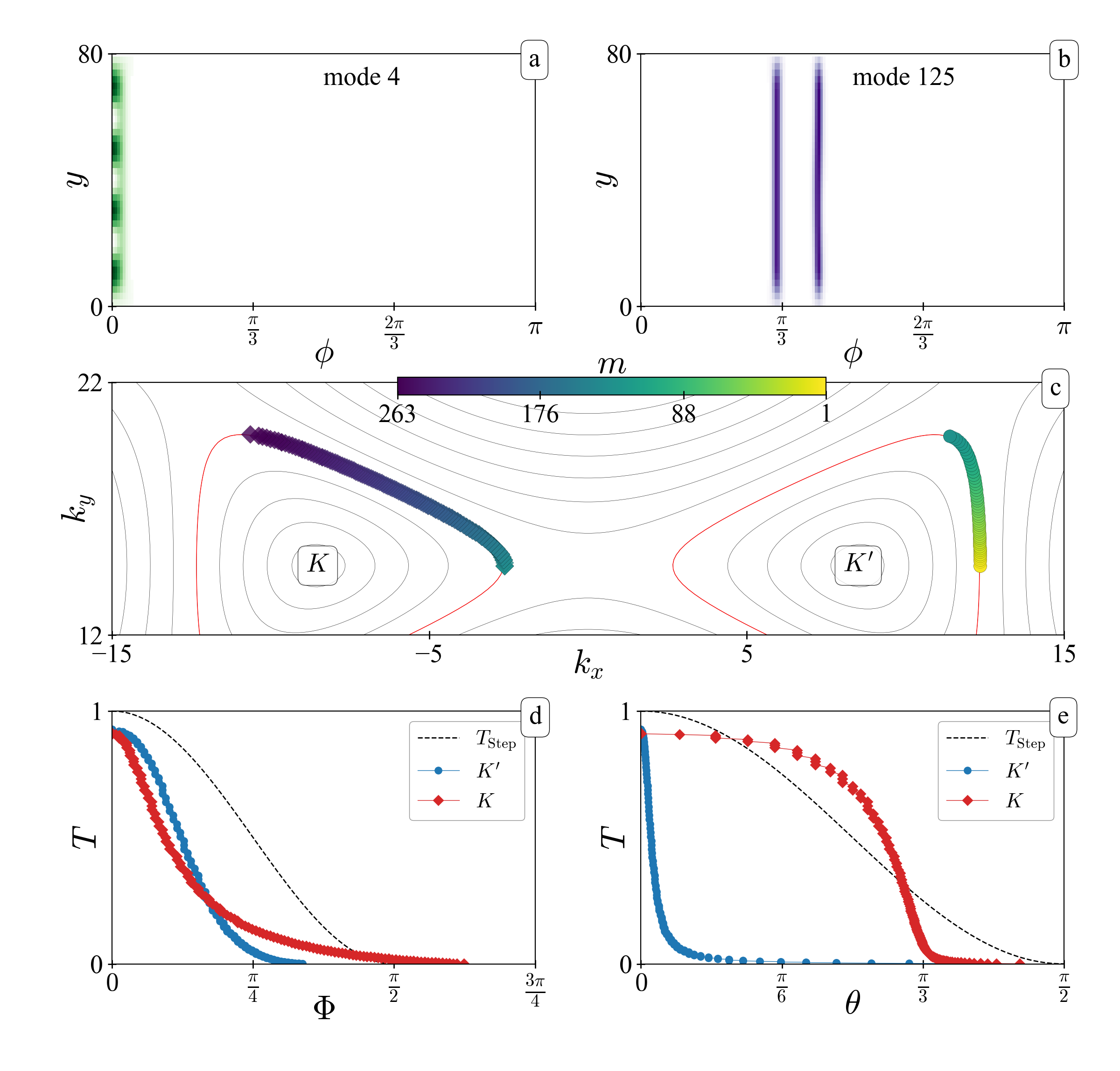}
	\caption{Transmission and Husimi functions in device A for high energies: $\sigma = 4, L = 12\sigma, W = 90, V_0 = 5, E_F = 0, w = 1$. (a, b) $Q(\phi, y)$ in $K_2'$ for two different modes. (c) Maxima of incoming $Q$ (see sec.~\ref{sec:higher_energy}) on momentum space. Each incoming mode $m$ is using a different color from the colorbar. (d) Mode transmission $T_m$ versus wavevector angle $\Phi_m$ (obtained using the Husimi function). (e) Same but versus group velocity angle $\theta_m$ instead. The dashed line plot of $T_\text{Step}$ is only meant as a guide to the eye, the formula is not valid for high energies.
	}
	\label{fig:high_energy}
\end{figure}

Klein tunneling applies to graphene because for small energies the Dirac equation is a valid approximation. In Klein tunneling the important angle is the wavevector angle (with respect to the Dirac points), see eqs.~\eqref{eq:Tstep}, \eqref{eq:TWKB}. The group velocity angle $\theta$ coincides with $\phi$ for small energies, however as the energy increases and trigonal warping effects begin to be significant, this is not the case anymore and $\theta \neq \phi$~\cite{DasSarma2011}. As there is no theoretical result on the tunneling behavior of graphene for energies beyond the Dirac regime, one is left to wonder: for higher energies is the Klein tunneling picture still relevant? And if yes, are the tunneling properties still dictated by $\phi$? This is an interesting question since the \emph{physical} propagation direction is governed by $\theta$.

We can answer this using the Husimi function. We significantly increase the energies in device A, setting $V_0 = 5$  and keeping $E_F = 0$, yielding incoming energy of $E = 2.5 \approx 0.9t$ which shows strong trigonal warping. Once again we compute incoming angles using $\Phi$ as in eq.~\eqref{eq:Phi} because incoming $Q$ is well-localized in momentum space, see fig.~\ref{fig:high_energy}a,b. However, the limits of $\text{argmax}$ must be modified. For modes $m \le \floor{M/2}$ the angle span of eq.~\eqref{eq:Phi} is set to $[0, \frac{\pi}{3})$ while for the rest of the modes it is set to $[0, \frac{2\pi}{3})$, due to the warping of the energy contour, see below.
This regime is not covered by~\cite{Brey2006a} and so the transverse wavefunctions are not necessarily sines (and we found $Q$ to perform much better than attempting to anyway fit sines to the transverse wavefunctions).

At higher energies the two valleys provide very different incoming modes, as seen in Fig.~\ref{fig:high_energy}c. For valley $K'$ there is a ``flat'' front, greatly limiting the possible group velocities. The contrary is happening in valley $K$ where the contour with positive group velocity spans more angles. In addition, in the $K'$ case the incoming wavevector angle is limited to $|\phi_\text{in}| \lesssim \pi/3$ but in $K$ we have $|\phi_\text{in}| \lesssim 2\pi/3$, due to the requirement of positive x-component of the group velocity, given by the divergence of the dispersion relation of graphene, eq.~\eqref{eq:dispersion}
\begin{align}
\begin{array}{rcl}
v_{x} &=& \frac{- \sqrt{3}\lambda t  a}{\sqrt{f(k) + 3}} \left(\sin\left(\frac{\sqrt{3} a}{2} k_{x}\right) \cos\left(\frac{3 a}{2} k_{y}\right) + \sin\left(\sqrt{3} k_{x} a\right)\right)\\
v_{y} &=& \frac{- 3\lambda a }{\sqrt{f(k) + 3}} \cos\left(\frac{\sqrt{3} a}{2} k_{x}\right)\sin\left(\frac{3 a}{2} k_{y}\right)
\end{array}
\label{eq:group}
\end{align}
($k_x,k_y$ are measured with respect to the center of the BZ here).

Klein tunneling assumes equivalence between the two valleys as it depends on the wavevector angle. To see whether some remnant of Klein tunneling exists at higher energies, we have to look for some tunneling property that not only decays exponentially with increasing angle of incidence, but also stays ``as similar'' as possible between the two valleys. In fig.~\ref{fig:high_energy}d,e we compare the transmission probability of each mode $T_m$ versus the wavevector angle $\Phi$ and group velocity angle $\theta$.

The result surprised us, since we find a Klein tunneling-like behaviour in $T_m$ versus $\phi$. We were rather expecting $T_m$ versus $\theta$ to  show similar behaviour at the two valleys, because $\theta$ corresponds to the physical propagation direction. We do not suggest that Klein tunneling straightforwardly applies to higher energies. In Fig.~\ref{fig:high_energy}f the characteristic perfect transmission at normal incidence ($\phi_\text{in} = 0$) is lost, nevertheless, it is clear that the tunneling probability as a function of the wavevector angle is quite similar to what would be expected for Klein tunneling.

\section{Applications of the Husimi function: device B}
\label{sec:deviceB}

In this section we study transport through the asymmetric device B (see Fig.~\ref{fig:devices}c) in which the incoming modes are scattered both from the boundary (``scattering edge'', highlighted in green) and the pn-junction. 
There are two main questions we want to address. First, to what extend can we use the existing expressions describing Klein tunneling to understand the transmission properties of such a device? These expressions are derived for plane waves, which have infinite spatial extend and are characterized by a single angle $\phi_\text{in}$.
Due to the boundary induced scattering the wavefunction in device B cannot be well approximated by a single plane wave. Can we use the Husimi technique to connect the transmission through the device to Klein tunneling? And also, how much can we push this technique, with respect to the physical size of the configurations we can examine?

Second, we want to understand how the type of the scattering edge affects intervalley scattering. There is strong theoretical evidence that the \emph{armchair} termination is in some way unique, while a random termination behaves like zigzag~\cite{McCann2004, Akhmerov2008, VanOstaay2011}. In addition, in the theoretical treatment of graphene nanoribbons in~\cite{Brey2006a}, the authors showed that the armchair termination \emph{mixes} valleys while the zigzag keeps them separated.
These (purely qualitative) arguments suggest that intervalley scattering should be enhanced by an edge with armchair termination.
Mason \textit{et al.} have shown in~\cite{Mason2013a} that a Husimi-based qualitative measure of intervalley scattering is generally enhanced at armchair boundaries. 
Here we quantify this effect by using the Husimi function, similarly as in~\ref{sec:intervalley} and we will show that intervalley scattering is indeed enhanced drastically at armchair edges.

\begin{figure*}[t!]
	\includegraphics[width=\textwidth]{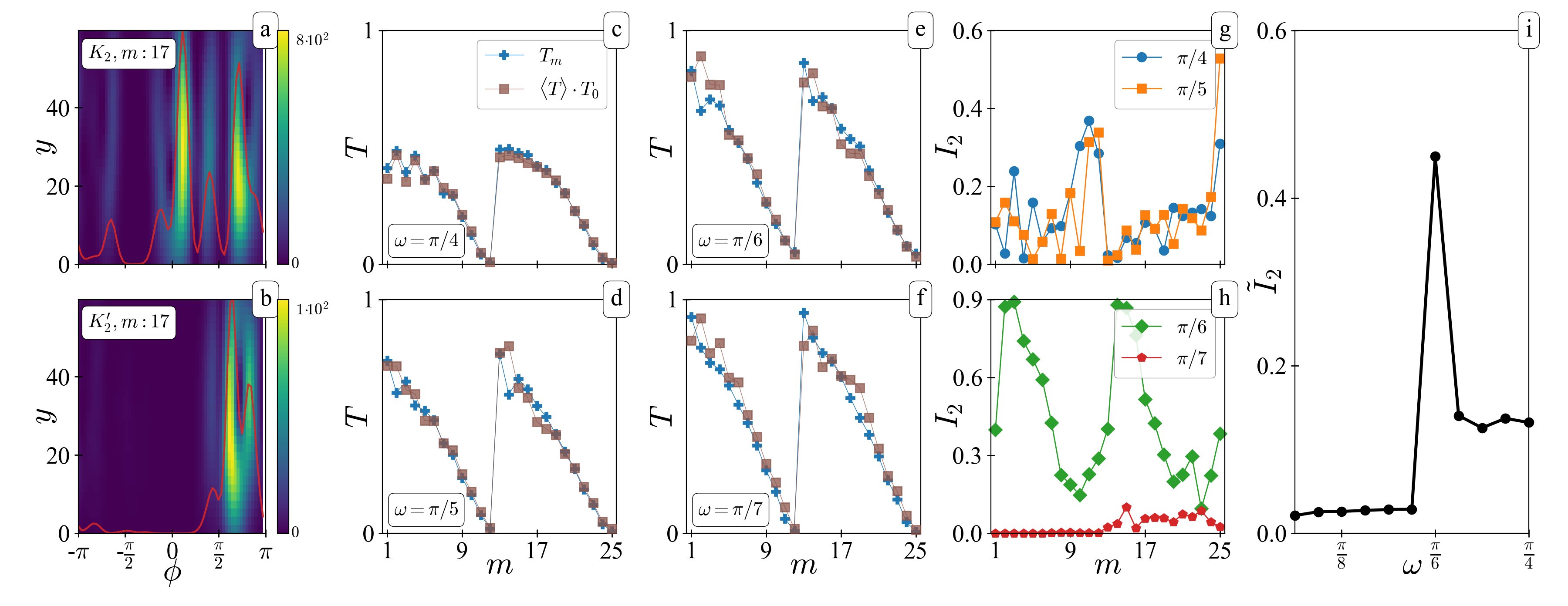}
	\caption{Tunneling and intervalley scattering in device B, using $\sigma = 10, L_1=L_2=6\sigma,W_1=120,w=1,V_0=0.4,E_F=0$. $W_2=W_1 - L_1\tan(\omega)$ depends on $\omega$. (a, b) Husimi function $Q(\phi,y)$ at position $x = L_1 + L_2/2$ (3$\sigma$ before the pn-junction) for $\omega=\pi/4$. For the mode shown, the incoming valley is $K_2$ (but a lot of intervalley scattering has already occurred). (c-f) Integrated transmission. (g, h) Intervalley scattering $I_2$ of eq.~\eqref{eq:intervalley} for various $\omega$ values using $Q$ measured at $x=L_1$ (computed without a pn-junction, $V_0 = 0, E_F =0.2$). (i) Average intervalley scattering per mode $\tilde{I}_2$ versus boundary angle $\omega$. A sharp increase is seen when $\omega = \pi/6$.}
	\label{fig:deviceB}
\end{figure*}

Let us stress that in device B the lead modes and the angles $\nu$ are not of much use, since the waves are deflected by the titled boundary of device B and also because $k_y$ is not conserved until the pn-junction.
On the other hand, $Q(\phi)$ is just as valid here as it was in sec.~\ref{sec:klein}. It also becomes clear from Fig.~\ref{fig:deviceB}a that many of the scattering waves inside $L_2$ cannot be approximated using a single angle, which means that one needs the entire distribution.

\subsection{Tunneling}
We find that we can apply the Klein tunneling formulas ``locally'' even in small devices and when the incoming waves are not single plane waves. We show this numerically using the integrated transmission formula, eq.~\eqref{eq:integrated} with $Q$ measured at location $x = L_1 + L_2/2$ (which is 3$\sigma$ before the pn-junction). However, now we can't compare $\langle T \rangle$ with $T_m$ directly, because $T_m$ also accounts for the back-scattering from the boundary inside $L_1$. To compensate for that, we compute the transmission of eq.~\eqref{eq:Tm} once without any pn-junction at all. We call this quantity $T_0$. We now have to compare $\langle T \rangle \cdot T_0$ with $T_m$, which we do in Fig.~\ref{fig:deviceB}c-f for various orientations of the boundary.

We see that the integrated transmission matches the transmission obtained through the pn-junction (using the scattering matrix) very well. This good agreement means that the Klein tunneling formula still locally describes the tunneling properties  at the pn-junction, even when the nanodevice is small (we found good results for $W_2$ as small as 20 nm) and the incoming wave is not a simple plane wave. In addition, this also means that the Husimi function accurately decomposed the incoming scattering wave into a representative distribution of angles of incidence. From this we can see that $Q$ allows us to separate the contributions of Klein tunneling (or any other transmission function $T(\ldots)$) in $T_m$, which could be useful in other scenarios as well.

\subsection{Intervalley Scattering}
We now want to explore the intervalley scattering induced by the scattering edge and not the pn-junction. Therefore we first obtain the scattering wavefunctions $\psi_m$ in device B without a pn-junction (i.e. $V_0=0, E_F=0.2$ eV). We measure $Q$ using a slice at $x=L_1$ (exactly where the scattering boundary ends) and we compute $I_2$ from $Q$ there. The results are shown in Fig.~\ref{fig:deviceB}g-i. 
We note that the results we display below do not have significant differences if one uses slices at $x>L_1$.

An important benefit of using $I_2$ (over e.g. measures used in~\cite{Mason2013a}) is that it does not depend on, or demands measuring $Q$ for $\textbf{r}_0$ \emph{exactly} at the boundaries of the nanodevice. This is crucial as the accuracy of the Husimi function dramatically drops at the boundaries, since most lattice sites around a circle of $3\sigma$ from $\mathbf{r}_0$ do not even exist. We have observed in our simulations that this leads to numeric artifacts and should be avoided, and we have also established this to be true in the analytic treatment of $Q$ in sec.~\ref{sec:lead_Q}.

There are two interesting observations to be made. First, the intervalley scattering from a lattice termination is fundamentally different from that seen in sec.~\ref{sec:intervalley} which results from a pn-junction. In the present case \emph{both} valleys always undergo intervalley scattering.

The second observation is what we expected from existing theory and now quantified using a well-defined measure: armchair lattice terminations induce much more intervalley scattering than any other termination orientation. This can be seen firstly in Fig.~\ref{fig:deviceB}g,h where $I_2$ has clearly higher values, but most prominently in panel i where we plot the average intervalley scattering per mode, i.e.
\begin{align}
\tilde{I}_2 = \frac{1}{M}\sum_{m=1}^{M} I_2(m).
\label{eq:I2tilde}
\end{align}
$\tilde{I}_2$ has a very sharp peak at $\omega = \pi/6$, where the boundary termination is exactly armchair.

\section{Conclusion}
In this paper we have used the Husimi function both numerically and analytically to analyse quantum transport through tight-binding nanodevices and have demonstrated that it is a very useful tool.
We have for example shown that even in situations where the angle of incidence on a tunnel barrier is not easily discernible we can use the Husimi distribution to evaluate Klein tunneling at this barrier. 
For higher Fermi energies the Husimi function allowed us to analyse the tunneling behavior in the regime of triangular warped Dirac cones.
We have also shown how $Q$ can be straightforwardly used to accurately define and measure intervalley scattering. 
Through this we have shown that pn-junctions not only introduce intervalley scattering, but that is unexpectedly strongly valley-asymmetric.
We also confirmed quantitatively that the strongest geometric intervalley scatterer in graphene is indeed the armchair termination.

A main goal in this work was to show that the Husimi function is a useful tool that solves the ``black box'' problem mentioned in the introduction and should be useful to have in the toolbox of condensed matter physicists.
The Husimi function \emph{complements}, and not competes with, the scattering matrix approach. 
Notice that to compute $Q$ numerically one needs the scattering wavefunctions.
What the Husimi function is able to do is to offer an additional level of depth that allows one to look directly into the device and even specific parts of the device (in both coordinate and momentum space at the same time, not possible just from the scattering wavefunctions $\psi_m$).
Whether this level of detail is necessary or useful depends of course on the exact problem one wants to study, and thus cannot be discussed generally.
What is important is that if such a level of detail is sought after, the Husimi function can provide it.
We also point out that even though the current work was applied to graphene, the methodology involving the Husimi function is in no way limited to it.

\textbf{Acknowledgments}
 The authors would like to thank Theo Geisel for helpful contributions.

\appendix

\section{Angle uncertainty}
\label{sec:angle_unc}
For a given value of the parameter $\sigma$, the wavepacket has a known uncertainty in both position and momentum
\begin{equation}
\sigma := \Delta x = \frac{1}{2\Delta k} .
\end{equation}
What we are interested about is the uncertainty in the propagation angle.
For small energies the propagation angle is the same for the wavevector and the group velocity defined as
 \begin{equation}
\phi = \arctan (q_y/q_x)
\end{equation} 
with $\mathbf{q} =\mathbf{k} - K^\xi$. For any nonlinear function, uncertainty propagation is given by
$$
\sigma_\phi^2 =  \left|\frac{\partial \phi}{\partial q_x} \sigma_{q_x}\right|^2 + \left|\frac{\partial \phi}{\partial q_y} \sigma_{q_y}\right|^2  .
$$
Since by definition $\sigma_{q_x}=\sigma_{q_y}= \Delta q = \Delta k$ we have
\begin{align*}
\sigma_\phi^2 & = \frac{q_y^2 \Delta q^2}{(q_x^2 + q_y^2)^2} + \frac{q_x^2 \Delta q^2}{(q_x^2 + q_y^2)^2} = \Delta q^2 \frac{q^2}{q^4} 
\end{align*} 
therefore we see that 
\begin{equation}
\sigma_\phi = \frac{\Delta q}{q}.
\end{equation} 
If we want to have a constant $\sigma_\phi$ for measurements at different energies, then we will use $\sigma$ such that (assuming also $\Delta x \Delta k = 1/2$)
\begin{equation}
\frac{\Delta q}{q} = \frac{1}{2q\sigma}\; \Rightarrow \;\sigma =  \frac{1}{2q} \left(\frac{\Delta q}{q}\right)^{-1}\;\Rightarrow \;\sigma = \frac{1}{2\sigma_\phi q}.
\end{equation}

\bibliographystyle{apsrev4-1} 
\bibliography{paper}

\begin{thebibliography}{45}%
\makeatletter
\providecommand \@ifxundefined [1]{%
 \@ifx{#1\undefined}
}%
\providecommand \@ifnum [1]{%
 \ifnum #1\expandafter \@firstoftwo
 \else \expandafter \@secondoftwo
 \fi
}%
\providecommand \@ifx [1]{%
 \ifx #1\expandafter \@firstoftwo
 \else \expandafter \@secondoftwo
 \fi
}%
\providecommand \natexlab [1]{#1}%
\providecommand \enquote  [1]{``#1''}%
\providecommand \bibnamefont  [1]{#1}%
\providecommand \bibfnamefont [1]{#1}%
\providecommand \citenamefont [1]{#1}%
\providecommand \href@noop [0]{\@secondoftwo}%
\providecommand \href [0]{\begingroup \@sanitize@url \@href}%
\providecommand \@href[1]{\@@startlink{#1}\@@href}%
\providecommand \@@href[1]{\endgroup#1\@@endlink}%
\providecommand \@sanitize@url [0]{\catcode `\\12\catcode `\$12\catcode
  `\&12\catcode `\#12\catcode `\^12\catcode `\_12\catcode `\%12\relax}%
\providecommand \@@startlink[1]{}%
\providecommand \@@endlink[0]{}%
\providecommand \url  [0]{\begingroup\@sanitize@url \@url }%
\providecommand \@url [1]{\endgroup\@href {#1}{\urlprefix }}%
\providecommand \urlprefix  [0]{URL }%
\providecommand \Eprint [0]{\href }%
\providecommand \doibase [0]{http://dx.doi.org/}%
\providecommand \selectlanguage [0]{\@gobble}%
\providecommand \bibinfo  [0]{\@secondoftwo}%
\providecommand \bibfield  [0]{\@secondoftwo}%
\providecommand \translation [1]{[#1]}%
\providecommand \BibitemOpen [0]{}%
\providecommand \bibitemStop [0]{}%
\providecommand \bibitemNoStop [0]{.\EOS\space}%
\providecommand \EOS [0]{\spacefactor3000\relax}%
\providecommand \BibitemShut  [1]{\csname bibitem#1\endcsname}%
\let\auto@bib@innerbib\@empty
\bibitem [{\citenamefont {Novoselov}\ \emph {et~al.}(2007)\citenamefont
  {Novoselov}, \citenamefont {Jiang}, \citenamefont {Zhang}, \citenamefont
  {Morozov}, \citenamefont {Stormer}, \citenamefont {Zeitler}, \citenamefont
  {Maan}, \citenamefont {Boebinger}, \citenamefont {Kim},\ and\ \citenamefont
  {Geim}}]{Novoselov2007}%
  \BibitemOpen
  \bibfield  {author} {\bibinfo {author} {\bibfnamefont {K.~S.}\ \bibnamefont
  {Novoselov}}, \bibinfo {author} {\bibfnamefont {Z.}~\bibnamefont {Jiang}},
  \bibinfo {author} {\bibfnamefont {Y.}~\bibnamefont {Zhang}}, \bibinfo
  {author} {\bibfnamefont {S.~V.}\ \bibnamefont {Morozov}}, \bibinfo {author}
  {\bibfnamefont {H.~L.}\ \bibnamefont {Stormer}}, \bibinfo {author}
  {\bibfnamefont {U.}~\bibnamefont {Zeitler}}, \bibinfo {author} {\bibfnamefont
  {J.~C.}\ \bibnamefont {Maan}}, \bibinfo {author} {\bibfnamefont {G.~S.}\
  \bibnamefont {Boebinger}}, \bibinfo {author} {\bibfnamefont {P.}~\bibnamefont
  {Kim}}, \ and\ \bibinfo {author} {\bibfnamefont {A.~K.}\ \bibnamefont
  {Geim}},\ }\href {\doibase 10.1126/science.1137201} {\bibfield  {journal}
  {\bibinfo  {journal} {Science}\ }\textbf {\bibinfo {volume} {315}},\ \bibinfo
  {pages} {1379} (\bibinfo {year} {2007})}\BibitemShut {NoStop}%
\bibitem [{\citenamefont {McCann}\ \emph {et~al.}(2006)\citenamefont {McCann},
  \citenamefont {Kechedzhi}, \citenamefont {Fal'ko}, \citenamefont {Suzuura},
  \citenamefont {Ando},\ and\ \citenamefont {Altshuler}}]{McCann2006}%
  \BibitemOpen
  \bibfield  {author} {\bibinfo {author} {\bibfnamefont {E.}~\bibnamefont
  {McCann}}, \bibinfo {author} {\bibfnamefont {K.}~\bibnamefont {Kechedzhi}},
  \bibinfo {author} {\bibfnamefont {V.~I.}\ \bibnamefont {Fal'ko}}, \bibinfo
  {author} {\bibfnamefont {H.}~\bibnamefont {Suzuura}}, \bibinfo {author}
  {\bibfnamefont {T.}~\bibnamefont {Ando}}, \ and\ \bibinfo {author}
  {\bibfnamefont {B.~L.}\ \bibnamefont {Altshuler}},\ }\href {\doibase
  10.1103/PhysRevLett.97.146805} {\bibfield  {journal} {\bibinfo  {journal}
  {Physical Review Letters}\ }\textbf {\bibinfo {volume} {97}},\ \bibinfo
  {pages} {146805} (\bibinfo {year} {2006})}\BibitemShut {NoStop}%
\bibitem [{\citenamefont {Katsnelson}\ \emph {et~al.}(2006)\citenamefont
  {Katsnelson}, \citenamefont {Novoselov},\ and\ \citenamefont
  {Geim}}]{Katsnelson2006}%
  \BibitemOpen
  \bibfield  {author} {\bibinfo {author} {\bibfnamefont {M.~I.}\ \bibnamefont
  {Katsnelson}}, \bibinfo {author} {\bibfnamefont {K.~S.}\ \bibnamefont
  {Novoselov}}, \ and\ \bibinfo {author} {\bibfnamefont {a.~K.}\ \bibnamefont
  {Geim}},\ }\href {\doibase 10.1038/nphys384} {\bibfield  {journal} {\bibinfo
  {journal} {Nature physics}\ }\textbf {\bibinfo {volume} {2}},\ \bibinfo
  {pages} {15} (\bibinfo {year} {2006})},\ \Eprint
  {http://arxiv.org/abs/0604323} {arXiv:0604323 [cond-mat]} \BibitemShut
  {NoStop}%
\bibitem [{\citenamefont {{Das Sarma}}\ \emph {et~al.}(2011)\citenamefont {{Das
  Sarma}}, \citenamefont {Adam}, \citenamefont {Hwang},\ and\ \citenamefont
  {Rossi}}]{DasSarma2011}%
  \BibitemOpen
  \bibfield  {author} {\bibinfo {author} {\bibfnamefont {S.}~\bibnamefont {{Das
  Sarma}}}, \bibinfo {author} {\bibfnamefont {S.}~\bibnamefont {Adam}},
  \bibinfo {author} {\bibfnamefont {E.~H.}\ \bibnamefont {Hwang}}, \ and\
  \bibinfo {author} {\bibfnamefont {E.}~\bibnamefont {Rossi}},\ }\href
  {\doibase 10.1103/RevModPhys.83.407} {\bibfield  {journal} {\bibinfo
  {journal} {Reviews of Modern Physics}\ }\textbf {\bibinfo {volume} {83}},\
  \bibinfo {pages} {407} (\bibinfo {year} {2011})},\ \Eprint
  {http://arxiv.org/abs/1003.4731} {arXiv:1003.4731} \BibitemShut {NoStop}%
\bibitem [{\citenamefont {Datta}(1995)}]{Datta1995}%
  \BibitemOpen
  \bibfield  {author} {\bibinfo {author} {\bibfnamefont {S.}~\bibnamefont
  {Datta}},\ }\href@noop {} {\emph {\bibinfo {title} {{Electronic Transport in
  Mesoscopic Systems}}}}\ (\bibinfo  {publisher} {Cambridge University press},\
  \bibinfo {year} {1995})\BibitemShut {NoStop}%
\bibitem [{\citenamefont {Husimi}(1940)}]{Husimi1940}%
  \BibitemOpen
  \bibfield  {author} {\bibinfo {author} {\bibfnamefont {K.}~\bibnamefont
  {Husimi}},\ }\href {\doibase JST.Journalarchive/ppmsj1919/22.264} {\bibfield
  {journal} {\bibinfo  {journal} {Proceedings of the Physico-Mathematical
  Society of Japan. 3rd Series}\ }\textbf {\bibinfo {volume} {22}},\ \bibinfo
  {pages} {264} (\bibinfo {year} {1940})}\BibitemShut {NoStop}%
\bibitem [{\citenamefont {Schleich}(2001)}]{Schleich2001}%
  \BibitemOpen
  \bibfield  {author} {\bibinfo {author} {\bibfnamefont {W.~P.}\ \bibnamefont
  {Schleich}},\ }\href@noop {} {\emph {\bibinfo {title} {{Quantum Optics in
  Phase Space}}}}\ (\bibinfo {year} {2001})\BibitemShut {NoStop}%
\bibitem [{\citenamefont {Moya-Cessa}\ \emph {et~al.}(2008)\citenamefont
  {Moya-Cessa}, \citenamefont {Moya-Cessa},\ and\ \citenamefont
  {Berriel-Valdos}}]{Moya-Cessa2008}%
  \BibitemOpen
  \bibfield  {author} {\bibinfo {author} {\bibfnamefont {H.}~\bibnamefont
  {Moya-Cessa}}, \bibinfo {author} {\bibfnamefont {J.~R.}\ \bibnamefont
  {Moya-Cessa}}, \ and\ \bibinfo {author} {\bibfnamefont {L.~R.}\ \bibnamefont
  {Berriel-Valdos}},\ }\href@noop {} {\bibfield  {journal} {\bibinfo  {journal}
  {Introduction to quantum optics}\ ,\ \bibinfo {pages} {31}} (\bibinfo {year}
  {2008})}\BibitemShut {NoStop}%
\bibitem [{\citenamefont {Virovlyansky}\ \emph {et~al.}(2012)\citenamefont
  {Virovlyansky}, \citenamefont {Makarov},\ and\ \citenamefont
  {Prants}}]{Virovlyansky2012}%
  \BibitemOpen
  \bibfield  {author} {\bibinfo {author} {\bibfnamefont {A.~L.}\ \bibnamefont
  {Virovlyansky}}, \bibinfo {author} {\bibfnamefont {D.~V.}\ \bibnamefont
  {Makarov}}, \ and\ \bibinfo {author} {\bibfnamefont {S.~V.}\ \bibnamefont
  {Prants}},\ }\href@noop {} {\ ,\ \bibinfo {pages} {30} (\bibinfo {year}
  {2012})}\BibitemShut {NoStop}%
\bibitem [{\citenamefont {Nonnenmacher}\ and\ \citenamefont
  {Voros}(1998)}]{Nonnenmacher1998}%
  \BibitemOpen
  \bibfield  {author} {\bibinfo {author} {\bibfnamefont {S.}~\bibnamefont
  {Nonnenmacher}}\ and\ \bibinfo {author} {\bibfnamefont {A.}~\bibnamefont
  {Voros}},\ }\href {\doibase 10.1023/A:1023080303171} {\bibfield  {journal}
  {\bibinfo  {journal} {Journal of Statistical Physics}\ }\textbf {\bibinfo
  {volume} {92}},\ \bibinfo {pages} {431} (\bibinfo {year} {1998})}\BibitemShut
  {NoStop}%
\bibitem [{\citenamefont {Bäcker}(2003)}]{Baecker2003}%
  \BibitemOpen
  \bibfield  {author} {\bibinfo {author} {\bibfnamefont {A.}~\bibnamefont
  {Bäcker}},\ }in\ \href@noop {} {\emph {\bibinfo {booktitle} {The
  {Mathematical} {Aspects} of {Quantum} {Maps}}}},\ \bibinfo {editor} {edited
  by\ \bibinfo {editor} {\bibfnamefont {M.~D.}\ \bibnamefont {Esposti}}\ and\
  \bibinfo {editor} {\bibfnamefont {S.}~\bibnamefont {Graffi}}}\ (\bibinfo
  {publisher} {Springer Berlin Heidelberg},\ \bibinfo {address} {Berlin,
  Heidelberg},\ \bibinfo {year} {2003})\ pp.\ \bibinfo {pages}
  {91--144}\BibitemShut {NoStop}%
\bibitem [{\citenamefont {Bäcker}\ \emph {et~al.}(2004)\citenamefont
  {Bäcker}, \citenamefont {Fürstberger},\ and\ \citenamefont
  {Schubert}}]{Baecker2004}%
  \BibitemOpen
  \bibfield  {author} {\bibinfo {author} {\bibfnamefont {A.}~\bibnamefont
  {Bäcker}}, \bibinfo {author} {\bibfnamefont {S.}~\bibnamefont
  {Fürstberger}}, \ and\ \bibinfo {author} {\bibfnamefont {R.}~\bibnamefont
  {Schubert}},\ }\href {\doibase 10.1103/PhysRevE.70.036204} {\bibfield
  {journal} {\bibinfo  {journal} {Physical Review E}\ }\textbf {\bibinfo
  {volume} {70}},\ \bibinfo {pages} {036204} (\bibinfo {year}
  {2004})}\BibitemShut {NoStop}%
\bibitem [{\citenamefont {Bäcker}\ \emph {et~al.}(2005)\citenamefont
  {Bäcker}, \citenamefont {Ketzmerick},\ and\ \citenamefont
  {Monastra}}]{Baecker2005}%
  \BibitemOpen
  \bibfield  {author} {\bibinfo {author} {\bibfnamefont {A.}~\bibnamefont
  {Bäcker}}, \bibinfo {author} {\bibfnamefont {R.}~\bibnamefont {Ketzmerick}},
  \ and\ \bibinfo {author} {\bibfnamefont {A.~G.}\ \bibnamefont {Monastra}},\
  }\href {\doibase 10.1103/PhysRevLett.94.054102} {\bibfield  {journal}
  {\bibinfo  {journal} {Physical Review Letters}\ }\textbf {\bibinfo {volume}
  {94}},\ \bibinfo {pages} {054102} (\bibinfo {year} {2005})}\BibitemShut
  {NoStop}%
\bibitem [{\citenamefont {Toscano}\ \emph {et~al.}(2008)\citenamefont
  {Toscano}, \citenamefont {Kenfack}, \citenamefont {Carvalho}, \citenamefont
  {Rost},\ and\ \citenamefont {Ozorio~de Almeida}}]{Toscano2008}%
  \BibitemOpen
  \bibfield  {author} {\bibinfo {author} {\bibfnamefont {F.}~\bibnamefont
  {Toscano}}, \bibinfo {author} {\bibfnamefont {A.}~\bibnamefont {Kenfack}},
  \bibinfo {author} {\bibfnamefont {A.~R.}\ \bibnamefont {Carvalho}}, \bibinfo
  {author} {\bibfnamefont {J.~M.}\ \bibnamefont {Rost}}, \ and\ \bibinfo
  {author} {\bibfnamefont {A.~M.}\ \bibnamefont {Ozorio~de Almeida}},\ }\href
  {\doibase 10.1098/rspa.2007.0263} {\bibfield  {journal} {\bibinfo  {journal}
  {Proceedings of the Royal Society A: Mathematical, Physical and Engineering
  Sciences}\ }\textbf {\bibinfo {volume} {464}},\ \bibinfo {pages} {1503}
  (\bibinfo {year} {2008})}\BibitemShut {NoStop}%
\bibitem [{\citenamefont {Schanz}\ \emph {et~al.}(2005)\citenamefont {Schanz},
  \citenamefont {Dittrich},\ and\ \citenamefont {Ketzmerick}}]{Schanz2005}%
  \BibitemOpen
  \bibfield  {author} {\bibinfo {author} {\bibfnamefont {H.}~\bibnamefont
  {Schanz}}, \bibinfo {author} {\bibfnamefont {T.}~\bibnamefont {Dittrich}}, \
  and\ \bibinfo {author} {\bibfnamefont {R.}~\bibnamefont {Ketzmerick}},\
  }\href {\doibase 10.1103/PhysRevE.71.026228} {\bibfield  {journal} {\bibinfo
  {journal} {Physical Review E}\ }\textbf {\bibinfo {volume} {71}},\ \bibinfo
  {pages} {026228} (\bibinfo {year} {2005})}\BibitemShut {NoStop}%
\bibitem [{\citenamefont {Hentschel}\ \emph {et~al.}(2003)\citenamefont
  {Hentschel}, \citenamefont {Schomerus},\ and\ \citenamefont
  {Schubert}}]{Hentschel2003}%
  \BibitemOpen
  \bibfield  {author} {\bibinfo {author} {\bibfnamefont {M.}~\bibnamefont
  {Hentschel}}, \bibinfo {author} {\bibfnamefont {H.}~\bibnamefont
  {Schomerus}}, \ and\ \bibinfo {author} {\bibfnamefont {R.}~\bibnamefont
  {Schubert}},\ }\href {\doibase 10.1209/epl/i2003-00421-1} {\bibfield
  {journal} {\bibinfo  {journal} {EPL (Europhysics Letters)}\ }\textbf
  {\bibinfo {volume} {62}},\ \bibinfo {pages} {636} (\bibinfo {year}
  {2003})}\BibitemShut {NoStop}%
\bibitem [{\citenamefont {Wiersig}\ and\ \citenamefont
  {Hentschel}(2008)}]{Wiersig2008}%
  \BibitemOpen
  \bibfield  {author} {\bibinfo {author} {\bibfnamefont {J.}~\bibnamefont
  {Wiersig}}\ and\ \bibinfo {author} {\bibfnamefont {M.}~\bibnamefont
  {Hentschel}},\ }\href {\doibase 10.1103/PhysRevLett.100.033901} {\bibfield
  {journal} {\bibinfo  {journal} {Physical Review Letters}\ }\textbf {\bibinfo
  {volume} {100}},\ \bibinfo {pages} {033901} (\bibinfo {year}
  {2008})}\BibitemShut {NoStop}%
\bibitem [{\citenamefont {Bäcker}\ \emph {et~al.}(2009)\citenamefont
  {Bäcker}, \citenamefont {Ketzmerick}, \citenamefont {Löck}, \citenamefont
  {Wiersig},\ and\ \citenamefont {Hentschel}}]{Baecker2009}%
  \BibitemOpen
  \bibfield  {author} {\bibinfo {author} {\bibfnamefont {A.}~\bibnamefont
  {Bäcker}}, \bibinfo {author} {\bibfnamefont {R.}~\bibnamefont {Ketzmerick}},
  \bibinfo {author} {\bibfnamefont {S.}~\bibnamefont {Löck}}, \bibinfo
  {author} {\bibfnamefont {J.}~\bibnamefont {Wiersig}}, \ and\ \bibinfo
  {author} {\bibfnamefont {M.}~\bibnamefont {Hentschel}},\ }\href {\doibase
  10.1103/PhysRevA.79.063804} {\bibfield  {journal} {\bibinfo  {journal}
  {Physical Review A}\ }\textbf {\bibinfo {volume} {79}},\ \bibinfo {pages}
  {063804} (\bibinfo {year} {2009})}\BibitemShut {NoStop}%
\bibitem [{\citenamefont {Feist}\ \emph {et~al.}(2006)\citenamefont {Feist},
  \citenamefont {Bäcker}, \citenamefont {Ketzmerick}, \citenamefont {Rotter},
  \citenamefont {Huckestein},\ and\ \citenamefont {Burgdörfer}}]{Feist2006}%
  \BibitemOpen
  \bibfield  {author} {\bibinfo {author} {\bibfnamefont {J.}~\bibnamefont
  {Feist}}, \bibinfo {author} {\bibfnamefont {A.}~\bibnamefont {Bäcker}},
  \bibinfo {author} {\bibfnamefont {R.}~\bibnamefont {Ketzmerick}}, \bibinfo
  {author} {\bibfnamefont {S.}~\bibnamefont {Rotter}}, \bibinfo {author}
  {\bibfnamefont {B.}~\bibnamefont {Huckestein}}, \ and\ \bibinfo {author}
  {\bibfnamefont {J.}~\bibnamefont {Burgdörfer}},\ }\href {\doibase
  10.1103/PhysRevLett.97.116804} {\bibfield  {journal} {\bibinfo  {journal}
  {Physical Review Letters}\ }\textbf {\bibinfo {volume} {97}},\ \bibinfo
  {pages} {116804} (\bibinfo {year} {2006})}\BibitemShut {NoStop}%
\bibitem [{\citenamefont {Mason}\ \emph
  {et~al.}(2013{\natexlab{a}})\citenamefont {Mason}, \citenamefont {Borunda},\
  and\ \citenamefont {Heller}}]{Mason2013a}%
  \BibitemOpen
  \bibfield  {author} {\bibinfo {author} {\bibfnamefont {D.~J.}\ \bibnamefont
  {Mason}}, \bibinfo {author} {\bibfnamefont {M.~F.}\ \bibnamefont {Borunda}},
  \ and\ \bibinfo {author} {\bibfnamefont {E.~J.}\ \bibnamefont {Heller}},\
  }\href {\doibase 10.1103/PhysRevB.88.165421} {\bibfield  {journal} {\bibinfo
  {journal} {Physical Review B - Condensed Matter and Materials Physics}\
  }\textbf {\bibinfo {volume} {88}},\ \bibinfo {pages} {1} (\bibinfo {year}
  {2013}{\natexlab{a}})}\BibitemShut {NoStop}%
\bibitem [{\citenamefont {Mason}\ \emph
  {et~al.}(2013{\natexlab{b}})\citenamefont {Mason}, \citenamefont {Borunda},\
  and\ \citenamefont {Heller}}]{Mason2013}%
  \BibitemOpen
  \bibfield  {author} {\bibinfo {author} {\bibfnamefont {D.}~\bibnamefont
  {Mason}}, \bibinfo {author} {\bibfnamefont {M.}~\bibnamefont {Borunda}}, \
  and\ \bibinfo {author} {\bibfnamefont {E.}~\bibnamefont {Heller}},\ }\href
  {\doibase 10.1209/0295-5075/102/60005} {\bibfield  {journal} {\bibinfo
  {journal} {EPL (Europhysics Letters)}\ }\textbf {\bibinfo {volume} {60005}},\
  \bibinfo {pages} {1} (\bibinfo {year} {2013}{\natexlab{b}})},\ \Eprint
  {http://arxiv.org/abs/arXiv:1205.0291v3} {arXiv:arXiv:1205.0291v3}
  \BibitemShut {NoStop}%
\bibitem [{\citenamefont {Mason}\ \emph {et~al.}(2015)\citenamefont {Mason},
  \citenamefont {Borunda},\ and\ \citenamefont {Heller}}]{Mason2015}%
  \BibitemOpen
  \bibfield  {author} {\bibinfo {author} {\bibfnamefont {D.~J.}\ \bibnamefont
  {Mason}}, \bibinfo {author} {\bibfnamefont {M.~F.}\ \bibnamefont {Borunda}},
  \ and\ \bibinfo {author} {\bibfnamefont {E.~J.}\ \bibnamefont {Heller}},\
  }\href {\doibase 10.1103/PhysRevB.91.165405} {\bibfield  {journal} {\bibinfo
  {journal} {Physical Review B}\ }\textbf {\bibinfo {volume} {91}},\ \bibinfo
  {pages} {165405} (\bibinfo {year} {2015})}\BibitemShut {NoStop}%
\bibitem [{\citenamefont {Klein}(1929)}]{Klein1929}%
  \BibitemOpen
  \bibfield  {author} {\bibinfo {author} {\bibfnamefont {O.}~\bibnamefont
  {Klein}},\ }\href {\doibase 10.1007/BF01339716} {\bibfield  {journal}
  {\bibinfo  {journal} {Zeitschrift f{\"{u}}r Physik}\ }\textbf {\bibinfo
  {volume} {53}},\ \bibinfo {pages} {157} (\bibinfo {year} {1929})}\BibitemShut
  {NoStop}%
\bibitem [{\citenamefont {Calogeracos}\ and\ \citenamefont
  {Dombey}(1999)}]{Calogeracos1999}%
  \BibitemOpen
  \bibfield  {author} {\bibinfo {author} {\bibfnamefont {A.}~\bibnamefont
  {Calogeracos}}\ and\ \bibinfo {author} {\bibfnamefont {N.}~\bibnamefont
  {Dombey}},\ }\href {\doibase 10.1080/001075199181387} {\bibfield  {journal}
  {\bibinfo  {journal} {Contemporary Physics}\ }\textbf {\bibinfo {volume}
  {40}},\ \bibinfo {pages} {313} (\bibinfo {year} {1999})}\BibitemShut
  {NoStop}%
\bibitem [{\citenamefont {Allain}\ and\ \citenamefont
  {Fuchs}(2011)}]{Allain2011}%
  \BibitemOpen
  \bibfield  {author} {\bibinfo {author} {\bibfnamefont {P.~E.}\ \bibnamefont
  {Allain}}\ and\ \bibinfo {author} {\bibfnamefont {J.~N.}\ \bibnamefont
  {Fuchs}},\ }\href {\doibase 10.1140/epjb/e2011-20351-3} {\bibfield  {journal}
  {\bibinfo  {journal} {European Physical Journal B}\ }\textbf {\bibinfo
  {volume} {83}},\ \bibinfo {pages} {301} (\bibinfo {year} {2011})},\ \Eprint
  {http://arxiv.org/abs/1104.5632} {arXiv:1104.5632} \BibitemShut {NoStop}%
\bibitem [{\citenamefont {Cheianov}\ and\ \citenamefont
  {Fal'ko}(2006)}]{Cheianov2006}%
  \BibitemOpen
  \bibfield  {author} {\bibinfo {author} {\bibfnamefont {V.~V.}\ \bibnamefont
  {Cheianov}}\ and\ \bibinfo {author} {\bibfnamefont {V.~I.}\ \bibnamefont
  {Fal'ko}},\ }\href {\doibase 10.1103/PhysRevB.74.041403} {\bibfield
  {journal} {\bibinfo  {journal} {Physical Review B}\ }\textbf {\bibinfo
  {volume} {74}},\ \bibinfo {pages} {041403} (\bibinfo {year} {2006})},\
  \Eprint {http://arxiv.org/abs/0603624} {arXiv:0603624 [cond-mat]}
  \BibitemShut {NoStop}%
\bibitem [{\citenamefont {Brey}\ and\ \citenamefont
  {Fertig}(2006)}]{Brey2006a}%
  \BibitemOpen
  \bibfield  {author} {\bibinfo {author} {\bibfnamefont {L.}~\bibnamefont
  {Brey}}\ and\ \bibinfo {author} {\bibfnamefont {H.~a.}\ \bibnamefont
  {Fertig}},\ }\href {\doibase 10.1103/PhysRevB.73.235411} {\bibfield
  {journal} {\bibinfo  {journal} {Physical Review B - Condensed Matter and
  Materials Physics}\ }\textbf {\bibinfo {volume} {73}},\ \bibinfo {pages} {2}
  (\bibinfo {year} {2006})},\ \Eprint {http://arxiv.org/abs/0603107}
  {arXiv:0603107 [cond-mat]} \BibitemShut {NoStop}%
\bibitem [{\citenamefont {Akhmerov}\ and\ \citenamefont
  {Beenakker}(2008)}]{Akhmerov2008}%
  \BibitemOpen
  \bibfield  {author} {\bibinfo {author} {\bibfnamefont {A.~R.}\ \bibnamefont
  {Akhmerov}}\ and\ \bibinfo {author} {\bibfnamefont {C.~W.~J.}\ \bibnamefont
  {Beenakker}},\ }\href {\doibase 10.1103/PhysRevB.77.085423} {\bibfield
  {journal} {\bibinfo  {journal} {Physical Review B}\ }\textbf {\bibinfo
  {volume} {77}},\ \bibinfo {pages} {085423} (\bibinfo {year}
  {2008})}\BibitemShut {NoStop}%
\bibitem [{\citenamefont {Groth}\ \emph {et~al.}(2014)\citenamefont {Groth},
  \citenamefont {Wimmer}, \citenamefont {Akhmerov},\ and\ \citenamefont
  {Waintal}}]{Groth2014}%
  \BibitemOpen
  \bibfield  {author} {\bibinfo {author} {\bibfnamefont {C.~W.}\ \bibnamefont
  {Groth}}, \bibinfo {author} {\bibfnamefont {M.}~\bibnamefont {Wimmer}},
  \bibinfo {author} {\bibfnamefont {A.~R.}\ \bibnamefont {Akhmerov}}, \ and\
  \bibinfo {author} {\bibfnamefont {X.}~\bibnamefont {Waintal}},\ }\href
  {\doibase 10.1088/1367-2630/16/6/063065} {\bibfield  {journal} {\bibinfo
  {journal} {New Journal of Physics}\ }\textbf {\bibinfo {volume} {16}},\
  \bibinfo {pages} {1} (\bibinfo {year} {2014})},\ \Eprint
  {http://arxiv.org/abs/1309.2926} {arXiv:1309.2926} \BibitemShut {NoStop}%
\bibitem [{Note1()}]{Note1}%
  \BibitemOpen
  \bibinfo {note} {We note that this is stationary, not a time-dependent
  description}\BibitemShut {NoStop}%
\bibitem [{\citenamefont {Takahashi}(1986)}]{Takahashi1986}%
  \BibitemOpen
  \bibfield  {author} {\bibinfo {author} {\bibfnamefont {K.}~\bibnamefont
  {Takahashi}},\ }\href {\doibase 10.1143/jpsj.55.762} {\bibfield  {journal}
  {\bibinfo  {journal} {Journal of the Physical Society of Japan}\ }\textbf
  {\bibinfo {volume} {55}},\ \bibinfo {pages} {762} (\bibinfo {year}
  {1986})}\BibitemShut {NoStop}%
\bibitem [{\citenamefont {Harriman}(1988)}]{Harriman1988}%
  \BibitemOpen
  \bibfield  {author} {\bibinfo {author} {\bibfnamefont {J.~E.}\ \bibnamefont
  {Harriman}},\ }\href {\doibase 10.1063/1.454477} {\bibfield  {journal}
  {\bibinfo  {journal} {The Journal of Chemical Physics}\ }\textbf {\bibinfo
  {volume} {88}},\ \bibinfo {pages} {6399} (\bibinfo {year}
  {1988})}\BibitemShut {NoStop}%
\bibitem [{\citenamefont {Heller}(2018)}]{Heller2018}%
  \BibitemOpen
  \bibfield  {author} {\bibinfo {author} {\bibfnamefont {E.~J.}\ \bibnamefont
  {Heller}},\ }\href
  {https://www.ebook.de/de/product/30176425/eric_j_heller_the_semiclassical_way_to_dynamics_and_spectroscopy.html}
  {\emph {\bibinfo {title} {The Semiclassical Way to Dynamics and
  Spectroscopy}}}\ (\bibinfo  {publisher} {Princeton University Press},\
  \bibinfo {year} {2018})\BibitemShut {NoStop}%
\bibitem [{Note2()}]{Note2}%
  \BibitemOpen
  \bibinfo {note} {For each $\protect \mathbf {r}_0$ we use only lattice sites
  that are within $|\protect \mathbf {r}_j - \protect \mathbf {r}_0|\leq
  3\sigma $, to reduce computation time. Notice the complex conjugation $\psi
  ^*$ in eqs.~\protect \textup {\hbox {\mathsurround \z@ \protect \normalfont
  (\ignorespaces \ref {eq:husimi_def_continuous}\unskip \@@italiccorr )}} and
  \protect \textup {\hbox {\mathsurround \z@ \protect \normalfont
  (\ignorespaces \ref {eq:husimi_def_discrete}\unskip \@@italiccorr )}}, which
  sometimes is omitted in the literature. While for closed systems with
  time-reversal symmetry it can be omitted, it is crucial for open systems,
  which we explore here, and for systems with broken time reversal symmetry
  like those in magnetic fields.}\BibitemShut {Stop}%
\bibitem [{\citenamefont {Mahmud}\ \emph {et~al.}(2005)\citenamefont {Mahmud},
  \citenamefont {Perry},\ and\ \citenamefont {Reinhardt}}]{Mahmud2005}%
  \BibitemOpen
  \bibfield  {author} {\bibinfo {author} {\bibfnamefont {K.~W.}\ \bibnamefont
  {Mahmud}}, \bibinfo {author} {\bibfnamefont {H.}~\bibnamefont {Perry}}, \
  and\ \bibinfo {author} {\bibfnamefont {W.~P.}\ \bibnamefont {Reinhardt}},\
  }\href {\doibase 10.1103/PhysRevA.71.023615} {\bibfield  {journal} {\bibinfo
  {journal} {Physical Review A}\ }\textbf {\bibinfo {volume} {71}},\ \bibinfo
  {pages} {023615} (\bibinfo {year} {2005})}\BibitemShut {NoStop}%
\bibitem [{\citenamefont {{Castro Neto}}\ \emph {et~al.}(2009)\citenamefont
  {{Castro Neto}}, \citenamefont {Guinea}, \citenamefont {Peres}, \citenamefont
  {Novoselov},\ and\ \citenamefont {Geim}}]{CastroNeto2009}%
  \BibitemOpen
  \bibfield  {author} {\bibinfo {author} {\bibfnamefont {a.~H.}\ \bibnamefont
  {{Castro Neto}}}, \bibinfo {author} {\bibfnamefont {F.}~\bibnamefont
  {Guinea}}, \bibinfo {author} {\bibfnamefont {N.~M.~R.}\ \bibnamefont
  {Peres}}, \bibinfo {author} {\bibfnamefont {K.~S.}\ \bibnamefont
  {Novoselov}}, \ and\ \bibinfo {author} {\bibfnamefont {a.~K.}\ \bibnamefont
  {Geim}},\ }\href {\doibase 10.1103/RevModPhys.81.109} {\bibfield  {journal}
  {\bibinfo  {journal} {Reviews of Modern Physics}\ }\textbf {\bibinfo {volume}
  {81}},\ \bibinfo {pages} {109} (\bibinfo {year} {2009})},\ \Eprint
  {http://arxiv.org/abs/0709.1163} {arXiv:0709.1163} \BibitemShut {NoStop}%
\bibitem [{\citenamefont {Goerbig}(2011)}]{Goerbig2011a}%
  \BibitemOpen
  \bibfield  {author} {\bibinfo {author} {\bibfnamefont {M.~O.}\ \bibnamefont
  {Goerbig}},\ }\href {\doibase 10.1103/RevModPhys.83.1193} {\bibfield
  {journal} {\bibinfo  {journal} {Rev. Mod. Phys.}\ }\textbf {\bibinfo {volume}
  {83}},\ \bibinfo {pages} {1193} (\bibinfo {year} {2011})},\ \Eprint
  {http://arxiv.org/abs/1004.3396} {arXiv:1004.3396} \BibitemShut {NoStop}%
\bibitem [{\citenamefont {Morpurgo}\ and\ \citenamefont
  {Guinea}(2006)}]{Morpurgo2006}%
  \BibitemOpen
  \bibfield  {author} {\bibinfo {author} {\bibfnamefont {A.~F.}\ \bibnamefont
  {Morpurgo}}\ and\ \bibinfo {author} {\bibfnamefont {F.}~\bibnamefont
  {Guinea}},\ }\href {\doibase 10.1103/PhysRevLett.97.196804} {\bibfield
  {journal} {\bibinfo  {journal} {Physical Review Letters}\ }\textbf {\bibinfo
  {volume} {97}},\ \bibinfo {pages} {1} (\bibinfo {year} {2006})},\ \Eprint
  {http://arxiv.org/abs/0603789} {arXiv:0603789 [cond-mat]} \BibitemShut
  {NoStop}%
\bibitem [{\citenamefont {Hilke}\ \emph {et~al.}(2014)\citenamefont {Hilke},
  \citenamefont {Massicotte}, \citenamefont {Whiteway},\ and\ \citenamefont
  {Yu}}]{Hilke2014}%
  \BibitemOpen
  \bibfield  {author} {\bibinfo {author} {\bibfnamefont {M.}~\bibnamefont
  {Hilke}}, \bibinfo {author} {\bibfnamefont {M.}~\bibnamefont {Massicotte}},
  \bibinfo {author} {\bibfnamefont {E.}~\bibnamefont {Whiteway}}, \ and\
  \bibinfo {author} {\bibfnamefont {V.}~\bibnamefont {Yu}},\ }\href {\doibase
  10.1155/2014/737296} {\bibfield  {journal} {\bibinfo  {journal}
  {TheScientificWorldJournal}\ }\textbf {\bibinfo {volume} {2014}},\ \bibinfo
  {pages} {737296} (\bibinfo {year} {2014})},\ \Eprint
  {http://arxiv.org/abs/arXiv:1212.5334v1} {arXiv:arXiv:1212.5334v1}
  \BibitemShut {NoStop}%
\bibitem [{\citenamefont {Yan}\ \emph {et~al.}(2016)\citenamefont {Yan},
  \citenamefont {Han}, \citenamefont {Jia}, \citenamefont {Niu}, \citenamefont
  {Cai}, \citenamefont {Yu},\ and\ \citenamefont {Wu}}]{Yan2016}%
  \BibitemOpen
  \bibfield  {author} {\bibinfo {author} {\bibfnamefont {B.}~\bibnamefont
  {Yan}}, \bibinfo {author} {\bibfnamefont {Q.}~\bibnamefont {Han}}, \bibinfo
  {author} {\bibfnamefont {Z.}~\bibnamefont {Jia}}, \bibinfo {author}
  {\bibfnamefont {J.}~\bibnamefont {Niu}}, \bibinfo {author} {\bibfnamefont
  {T.}~\bibnamefont {Cai}}, \bibinfo {author} {\bibfnamefont {D.}~\bibnamefont
  {Yu}}, \ and\ \bibinfo {author} {\bibfnamefont {X.}~\bibnamefont {Wu}},\
  }\href {\doibase 10.1103/PhysRevB.93.041407} {\bibfield  {journal} {\bibinfo
  {journal} {Physical Review B}\ }\textbf {\bibinfo {volume} {93}},\ \bibinfo
  {pages} {1} (\bibinfo {year} {2016})},\ \Eprint
  {http://arxiv.org/abs/1602.06519} {arXiv:1602.06519} \BibitemShut {NoStop}%
\bibitem [{\citenamefont {Rycerz}\ \emph {et~al.}(2007)\citenamefont {Rycerz},
  \citenamefont {Tworzyd{\l}o},\ and\ \citenamefont {Beenakker}}]{Rycerz2007}%
  \BibitemOpen
  \bibfield  {author} {\bibinfo {author} {\bibfnamefont {A.}~\bibnamefont
  {Rycerz}}, \bibinfo {author} {\bibfnamefont {J.}~\bibnamefont
  {Tworzyd{\l}o}}, \ and\ \bibinfo {author} {\bibfnamefont {C.~W.}\
  \bibnamefont {Beenakker}},\ }\href {\doibase 10.1038/nphys547} {\bibfield
  {journal} {\bibinfo  {journal} {Nature Physics}\ }\textbf {\bibinfo {volume}
  {3}},\ \bibinfo {pages} {172} (\bibinfo {year} {2007})},\ \Eprint
  {http://arxiv.org/abs/0608533} {arXiv:0608533 [cond-mat]} \BibitemShut
  {NoStop}%
\bibitem [{\citenamefont {Gunlycke}\ and\ \citenamefont
  {White}(2011)}]{Gunlycke2011}%
  \BibitemOpen
  \bibfield  {author} {\bibinfo {author} {\bibfnamefont {D.}~\bibnamefont
  {Gunlycke}}\ and\ \bibinfo {author} {\bibfnamefont {C.~T.}\ \bibnamefont
  {White}},\ }\href {\doibase 10.1103/PhysRevLett.106.136806} {\bibfield
  {journal} {\bibinfo  {journal} {Physical Review Letters}\ }\textbf {\bibinfo
  {volume} {106}},\ \bibinfo {pages} {1} (\bibinfo {year} {2011})},\ \Eprint
  {http://arxiv.org/abs/1103.4313} {arXiv:1103.4313} \BibitemShut {NoStop}%
\bibitem [{\citenamefont {Lee}\ and\ \citenamefont {Lee}(2017)}]{Lee2017}%
  \BibitemOpen
  \bibfield  {author} {\bibinfo {author} {\bibfnamefont {K.~W.}\ \bibnamefont
  {Lee}}\ and\ \bibinfo {author} {\bibfnamefont {C.~E.}\ \bibnamefont {Lee}},\
  }\href {\doibase 10.1103/PhysRevB.95.195132} {\bibfield  {journal} {\bibinfo
  {journal} {Physical Review B}\ }\textbf {\bibinfo {volume} {95}},\ \bibinfo
  {pages} {1} (\bibinfo {year} {2017})}\BibitemShut {NoStop}%
\bibitem [{\citenamefont {McCann}\ and\ \citenamefont
  {Fal~ko}(2004)}]{McCann2004}%
  \BibitemOpen
  \bibfield  {author} {\bibinfo {author} {\bibfnamefont {E.}~\bibnamefont
  {McCann}}\ and\ \bibinfo {author} {\bibfnamefont {V.~I.}\ \bibnamefont
  {Fal~ko}},\ }\href {\doibase 10.1088/0953-8984/16/13/016} {\bibfield
  {journal} {\bibinfo  {journal} {Journal of Physics: Condensed Matter}\
  }\textbf {\bibinfo {volume} {16}},\ \bibinfo {pages} {2371} (\bibinfo {year}
  {2004})}\BibitemShut {NoStop}%
\bibitem [{\citenamefont {{Van Ostaay}}\ \emph {et~al.}(2011)\citenamefont
  {{Van Ostaay}}, \citenamefont {Akhmerov}, \citenamefont {Beenakker},\ and\
  \citenamefont {Wimmer}}]{VanOstaay2011}%
  \BibitemOpen
  \bibfield  {author} {\bibinfo {author} {\bibfnamefont {J.~A.}\ \bibnamefont
  {{Van Ostaay}}}, \bibinfo {author} {\bibfnamefont {A.~R.}\ \bibnamefont
  {Akhmerov}}, \bibinfo {author} {\bibfnamefont {C.~W.}\ \bibnamefont
  {Beenakker}}, \ and\ \bibinfo {author} {\bibfnamefont {M.}~\bibnamefont
  {Wimmer}},\ }\href {\doibase 10.1103/PhysRevB.84.195434} {\bibfield
  {journal} {\bibinfo  {journal} {Physical Review B - Condensed Matter and
  Materials Physics}\ }\textbf {\bibinfo {volume} {84}},\ \bibinfo {pages} {1}
  (\bibinfo {year} {2011})},\ \Eprint {http://arxiv.org/abs/1109.0884}
  {arXiv:1109.0884} \BibitemShut {NoStop}%
\end{thebibliography}%
	
\end{document}